\RequirePackage{fix-cm}
\documentclass[smallextended]{svjour3}       
\smartqed  
\usepackage[latin1]{inputenc}
\usepackage{amsmath}
\usepackage[ngerman,english]{babel}
\usepackage{amsfonts}
\usepackage{amssymb}
\usepackage{mathtools}
\usepackage{graphicx}
\usepackage[bookmarks]{hyperref}
\usepackage{dsfont}
\newcommand{\wt}[1]{\widetilde{#1}}
\newcommand{\deriv}[2]{\frac{d\, #1}{d #2}}
\newcommand{\parderiv}[2]{\frac{\partial\, #1}{\partial #2}}
\newcommand{\parderivTwo}[3]{\frac{\partial\,^2\, #1}{\partial #2\,\partial #3}}
\newcommand{\funcint}[1]{\int\mathcal{D}[#1]\,}
\newcommand{\funcderiv}[2]{\frac{\delta #1}{\delta #2}}

\newcommand{\expval}[1]{\left\langle #1\right\rangle}

\newcommand{\cc}[1]{\overline{#1}}
\newcommand{\inZ}[1]{#1\in\mathds{Z}}
\newcommand{\inZwo}[1]{#1\in\mathds{Z}\setminus\{0\}}
\newcommand{\inN}[1]{#1\in\mathds{N}}
\newcommand{\inR}[1]{#1\in\mathfrak{R}}
\newcommand{\inRwo}[1]{#1\in\mathfrak{R}\setminus\{0\}}
\newcommand{\inRWO}[2]{#1\in\mathfrak{R}\setminus\{0 #2\}}
\newcommand{\inRkWO}[3]{#1\in\mathfrak{R}_{#2}\setminus\{0 #3\}}
\newcommand{\hh}[2]{h_{#1}^{(#2)}}
\newcommand{\norm}[2]{\left\Vert#1\right\Vert_{\mathcal{L}_2 #2}}
\newcommand{\normZero}[1]{\left\Vert#1\right\Vert_0}
\newcommand{\fr}{\mathbf{r}}
\newcommand{\fPhi}{\mathbf{\Phi}}
\newcommand{\fj}{\mathbf{j}}
\allowdisplaybreaks

\begin{document}

\title{Field--Theoretic Thermodynamic Uncertainty Relation}
\subtitle{General formulation exemplified with the Kardar--Parisi--Zhang equation}

\author{Oliver Niggemann  \and  Udo Seifert}

\institute{O. Niggemann \at
              II. Institute for Theoretical Physics, University of Stuttgart, Pfaffenwaldring 57, 70550 Stuttgart, Germany \\
              \email{niggemann@theo2.physik.uni-stuttgart.de}
           \and
           U. Seifert \at
              II. Institute for Theoretical Physics, University of Stuttgart, Pfaffenwaldring 57, 70550 Stuttgart, Germany\\
              \email{useifert@theo2.physik.uni-stuttgart.de}
}

\date{Received: date / Accepted: date}

\maketitle

\begin{abstract}
We introduce a field-theoretic thermodynamic uncertainty relation as an extension of the one derived so far for a Markovian dynamics on a discrete set of states and for overdamped Langevin equations. We first formulate a framework which describes quantities like current, entropy production and diffusivity in the case of a generic field theory. We will then apply this general setting to the one-dimensional Kardar-Parisi-Zhang equation, a paradigmatic example of a non-linear field-theoretic Langevin equation. In particular, we will treat the dimensionless Kardar-Parisi-Zhang equation with an effective coupling parameter measuring the strength of the non-linearity. It will be shown that the field-theoretic thermodynamic uncertainty relation holds up to second order in a perturbation expansion with respect to a small effective coupling constant.
\end{abstract}

\section{Introduction}\label{sec:Intro}

The thermodynamic uncertainty relation (TUR) in a non-equilibrium steady state (NESS) provides a bound on the entropy production in terms of mean and variance of an arbitrary current \cite{BaratoSeifertUR2015}. Specifically, in the NESS, after a time $t$ a fluctuating integrated current $X(t)$ has a mean $\expval{X(t)}=j\,t$, and a diffusivity $D=\lim_{t\to\infty}\expval{(X(t)-j\,t)^2}/(2\,t)$. With the entropy production rate $\sigma$ the expectation of the total entropy production in the NESS is given by $\sigma\,t$. These quantities satisfy the universal thermodynamic uncertainty relation 
\begin{equation*}
\sigma\geq\frac{j^2}{D},
\end{equation*}
i.e. $\sigma$ is bounded from below by $j^2/D$. The TUR has been proven for a Markovian dynamics on a general network by Gingrich \textit{et al.} \cite{Gingrich2016,Horowitz2017} and further investigated for a number of different settings, both in the classical (see, e.g., \cite{Pietzonka2016,Polettini2016,Pietzonka2017,Proesmans2017,Gingrich2017,Gingrich2017FirstPassage,Garrahan2017,Dechant2018,Dechant20181,Koyuk2019,Chun2019,Falasco2019}) and the quantum domain (see, e.g., \cite{Brandner2018,Macieszczak2018,Agarwalla2018,Ptaszynski2018,Carrega2019,Guarnieri2019,Carollo2019}). It has led to a deeper understanding of systems far from equilibrium as it introduces a lower bound on the dissipation given the knowledge of the occurring fluctuations. Such a relation is of interest for the modeling and analysis of e.g. biomolecular processes, which may often be described as a Markov network (see e.g. \cite{Nguyen2016,Pietzonka2016Motor,Hwang2018}).\\
Of particular interest is the work by Gingrich \textit{et al.} \cite{Gingrich2017}, where the authors extend the relation from mesoscopic Markov jump processes to overdamped Langevin equations. Here a temporal coarse-graining procedure is described, which allows the formulation of a discrete Markov jump process in terms of an overdamped Langevin equation for the mesoscopic states of the model. These authors observe that for purely dissipative dynamics the TUR is saturated.  An additional spatial coarse-graining performed in \cite{Gingrich2017} results in a macroscopic description, where it is found that the tightness of the resulting uncertainty relation increases with the strength of the Gaussian potential wells (see \cite{Gingrich2017}, fig. $9$).\\
In this work, we present a field-theoretic equivalent to the TUR. Such a thermodynamic uncertainty relation for general field-theoretic Langevin equations may prove helpful in further understanding complex dynamics like turbulence for fluid flow or non-linear growth processes, described by the stochastic Navier-Stokes equation (e.g. \cite{FoiasBookNavierStokes2001}) or the Kardar-Parisi-Zhang equation \cite{KardarParisiZhang1986}, respectively. Both are prominent representatives of field-theoretic Langevin equations. For the latter, we highlight the recent progress concerning a study of the inward growth of interfaces in liquid crystal turbulence as an experimental realization. On the theory side,  analytic results on the effect of aging of two-time correlation functions for the interface growth were found \cite{NardisTakeuchi2017}. Furthermore we refer the reader to three review articles \cite{HalpinHealyTakeuchi2015,Spohn2017,Takeuchi2017} concerning the latest developments around the Kardar-Parisi-Zhang universality class.\\
The paper is organized as follows. In order to state a field-theoretic version of the thermodynamic uncertainty relation, we translate in \autoref{sec:TURFieldTheory} the notion of current, diffusivity and entropy production known from the setting of coupled Langevin equations to their respective equivalents for general field-theoretic Langevin equations. As an illustration of the generalizations introduced in \autoref{sec:TURFieldTheory}, we will then study the one-dimensional Kardar-Parisi-Zhang (KPZ) equation as a paradigmatic example of such a field-theoretic Langevin equation. As the calculation of the current, diffusivity and entropy production in the NESS requires a solution to the KPZ equation, we will use spectral theory and construct an approximate solution in the weak-coupling regime of the KPZ equation in \autoref{sec:TheoBack}. With this approximation, we will then derive in \autoref{sec:ThermoUncertRel} the thermodynamic uncertainty relation to quadratic order in the coupling parameter.

\section{Thermodynamic Uncertainty Relation for a Field Theory}\label{sec:TURFieldTheory}

In this section, we will present a generalization of the thermodynamic uncertainty relation introduced in \cite{BaratoSeifertUR2015} to a field theory. Consider a generic field theory of the form
\begin{align}
\begin{split}
\partial_t\Phi_\gamma(\fr,t)&=F_\gamma\left[\{\Phi_\mu(\fr,t)\}\right]+\eta_\gamma(\fr,t),\\
\expval{\eta_\gamma(\fr,t)}&=0,\\
\expval{\eta_\gamma(\fr,t)\eta_\kappa(\fr^\prime,t^\prime)}&=K(\fr-\fr^\prime)\delta_{\gamma,\,\kappa}\delta(t-t^\prime).
\end{split}\label{eq:GeneralFTLangevin}
\end{align}
Here $\Phi_\gamma(\fr,t)$ is a scalar field or the $\gamma$-th component of a vector field $(\gamma\in[1,n];\,n\in\mathds{N})$ with $\fr\in\Omega\subset\mathds{R}^d$, $F_\gamma\left[\{\Phi_\mu(\fr,t)\}\right]$ represents a (possibly non-linear) functional of $\Phi_\mu$ and $\eta_\gamma(\fr,t)$ denotes Gaussian noise, which is white in time, and with $K(\fr-\fr^\prime)$ as spatial noise correlations. Prominent examples of \eqref{eq:GeneralFTLangevin} are the stochastic Navier-Stokes equation for turbulent flow (see e.g. \cite{FoiasBookNavierStokes2001}) or the Kardar-Parisi-Zhang equation for non-linear growth processes \cite{KardarParisiZhang1986} to name only two. The latter will be treated in the subsequent sections within the framework established in the following.\par
Let us begin with the introduction of some notions. A natural choice of a local fluctuating current $\mathbf{j}(\fr,t)$ is 
\begin{equation}
\fj(\fr,t)\equiv\partial_t\fPhi(\fr,t),\label{eq:DefGenLocalCurrent}
\end{equation}
with $\fPhi(\fr,t)=\left(\Phi_1(\fr,t),\ldots,\Phi_n(\fr,t)\right)^\top$. The local current $\fj(\fr,t)$ is fluctuating around its mean, i.e.
\begin{equation}
\fj(\fr,t)=\expval{\fj(\fr,t)}+\delta\fj(\fr,t),\label{eq:DefGenLocalCurrentFluctuations}
\end{equation}
with $\delta\fj(\fr,t)$ denoting the fluctuations. Given that the system \eqref{eq:GeneralFTLangevin} possesses a NESS, the long-time behavior of the local current \eqref{eq:DefGenLocalCurrent} can be described as
\begin{equation}
\fj(\fr,t)=\mathbf{J}(\fr)+\delta\fj(\fr,t),\label{eq:DefGenLocalCurrentSteadyState}
\end{equation}
with
\begin{equation}
\mathbf{J}(\fr)=\lim_{t\to\infty}\partial_t\expval{\fPhi(\fr,t)}=\lim_{t\to\infty}\frac{\expval{\fPhi(\fr,t)}}{t},\label{eq:DefGenLocalStatCurrent}
\end{equation}
where $\expval{\cdot}$ denotes averages with respect to the noise history. As the thermodynamic uncertainty relation in a Markovian network is formulated for some form of integrated currents, we define in analogy the projection of the local current onto an arbitrarily directed weight function $\mathbf{g}(\fr)$
\begin{equation}
j_g(t)\equiv\int_\Omega d\fr\,\fj(\fr,t)\cdot\mathbf{g}(\fr).\label{eq:DefGenProjectedCurrentJg}
\end{equation}
The integral in \eqref{eq:DefGenProjectedCurrentJg} represents the usual $\mathcal{L}_2$-product of the two vector fields $\fj(\fr,t)$ and $\mathbf{g}(\fr)$ with $\fj(\fr,t)\cdot\mathbf{g}(\fr)=\sum_kj_k(\fr,t)g_k(\fr)$ as the scalar product between $\fj$ and $\mathbf{g}$. With this projected current $j_g(t)$, we associate a fluctuating \lq output\rq{}
\begin{equation}
\Psi_g(t)\equiv\int_\Omega d\fr\,\fPhi(\fr,t)\cdot\mathbf{g}(\fr).\label{eq:DefGenProjectedOutputPsig}
\end{equation}
Hence $j_g(t)=\partial_t\Psi_g(t)$ and in the NESS
\begin{equation}
J_g\equiv\lim_{t\to\infty}\frac{\expval{\Psi_g(t)}}{t}.\label{eq:DefGenProjectedCurrentSteadyState}
\end{equation}
The fluctuating output $\Psi_g(t)$ provides us with the means to define a measure of the precision of the system output, namely the squared variational coefficient $\epsilon^2$, as
\begin{equation}
\epsilon^2\equiv\frac{\expval{\left(\Psi_g(t)-\expval{\Psi_g(t)}\right)^2}}{\expval{\Psi_g(t)}^2}.\label{eq:DefGenEpsilonSquared}
\end{equation}
If the system is in its non-equilibrium steady state, we can rewrite \eqref{eq:DefGenEpsilonSquared} as
\begin{equation}
\epsilon^2=\frac{\expval{\left(\Psi_g(t)-J_g\,t\right)^2}}{\left(J_g\,t\right)^2}.\label{eq:DefGenEpsilonSquaredSteadyState}
\end{equation}
Let us now connect the variance of the output $\Psi_g(t)$ to the Green-Kubo diffusivity given by
\begin{equation}
D_g\equiv\int_0^\infty dt\,\expval{\delta j_g(t)\,\delta j_g(0)}.\label{eq:DefGenGreenKubo}
\end{equation}
Using \eqref{eq:DefGenProjectedCurrentJg} and \eqref{eq:DefGenLocalCurrent}, it is straightforward to verify that
\begin{align*}
\int_0^t dt^\prime\,\delta j_g(t^\prime)=\wt{\Psi}_g(t)-\expval{\wt{\Psi}_g(t)},\qquad\wt{\Psi}_g(t)\equiv\Psi_g(t)-\Psi_g(0).
\end{align*}
Thus,
\begin{equation}
\expval{\left(\wt{\Psi}_g(t)-\expval{\wt{\Psi}_g(t)}\right)^2}=\int_0^tdr\int_0^tds\,\expval{\delta j_g(r)\delta j_g(s)}.\label{eq:DefGenFiniteTimeGreenKubo}
\end{equation}
By dividing both sides of \eqref{eq:DefGenFiniteTimeGreenKubo} by $2t$ and taking the limit of $t\to\infty$ it is found in analogy to \cite{Kubo1966}, that
\begin{align*}
\lim_{t\to\infty}\frac{\int_0^tdr\int_0^tds\,\expval{\delta j_g(r)\delta j_g(s)}}{2t}=D_g,
\end{align*}
with $D_g$ from \eqref{eq:DefGenGreenKubo} and therefore
\begin{equation}
D_g=\lim_{t\to\infty}\frac{\expval{\left(\wt{\Psi}_g(t)-\expval{\wt{\Psi}_g(t)}\right)^2}}{2t}.\label{eq:DefGenDiffusivity}
\end{equation}
Since in the NESS $\Psi_g(t)$ is stochastically independent of the initial configuration $\Psi_g(0)$, we can simplify the expression for the diffusivity in the NESS according to
\begin{equation}
D_g=\lim_{t\to\infty}\frac{\expval{\left(\Psi_g(t)-\expval{\Psi_g(t)}\right)^2}}{2t}.\label{eq:DefGenNESSDiffusivity}
\end{equation}
With the result of \eqref{eq:DefGenNESSDiffusivity} and $\epsilon^2$ from \eqref{eq:DefGenEpsilonSquared}, an alternative formulation of the precision in a NESS is
\begin{equation}
\epsilon^2=\frac{\expval{\left(\Psi_g(t)-\expval{\Psi_g(t)}\right)^2}}{\expval{\Psi_g(t)}^2}=2\frac{D_g}{J_g^2}\frac{1}{t}.\label{eq:DefGenPrecisionViaDiffusivity}
\end{equation}
We proceed with expressing the total entropy production $\Delta s_\text{tot}$. The total entropy production is given by the sum of the entropy dissipated into the medium along a single trajectory, $\Delta s_\text{m}$, and the stochastic entropy, $\Delta s$, of such a trajectory; see e.g. \cite{SeifertReview2012}. The medium entropy is given by,
\begin{equation}
\Delta s_\text{m}\equiv\ln\frac{p[\fPhi(\fr,t)|\fPhi(\fr,t_0)]}{p[\wt{\fPhi}(\fr,t)|\wt{\fPhi}(\fr,t_0)]}.\label{eq:DefGenMediumEntropy}
\end{equation}
Here $p[\fPhi(\fr,t)|\fPhi(\fr,t_0)]$ denotes the functional probability density of the entire vector field $\fPhi(\fr,t)$, i.e. the field configuration after some time $t$ has elapsed since a starting-time $t_0<t$, conditioned on an initial value $\fPhi(\fr,t_0)$, i.e. a certain field configuration at the starting time $t_0$. In contrast, $p[\wt{\fPhi}(\fr,t)|\wt{\fPhi}(\fr,t_0)]$ is the conditioned probability density of the time reversed process, i.e. starting in the final configuration at time $t_0$ and ending up in the original one at time $t$. For the sake of simplicity, we will write in the following $p[\fPhi]$ and $p[\wt{\fPhi}]$ instead of $p[\fPhi(\fr,t)|\fPhi(\fr,t_0)]$ and $p[\wt{\fPhi}(\fr,t)|\wt{\fPhi}(\fr,t_0)]$, respectively. The functional probability density can be expressed via a so-called action functional, $\mathcal{S}[\fPhi]$, according to
\begin{equation}
p[\fPhi]\propto\exp\left[-\mathcal{S}[\fPhi]\right].\label{eq:DefGenProbDensity}
\end{equation}
For the system \eqref{eq:GeneralFTLangevin}, the action functional (see e.g. \cite{SeifertReview2012,Janssen1976,DeDominicis1976,MSR1973,NiggemannSincNoise2018,Hochberg2000,TaeuberBook2014,AltlandBook2010} and references therein) is given by 
\begin{align}
\begin{split}
\mathcal{S}[\fPhi]&=\frac{1}{2}\sum_\gamma\int_{t_0}^tdt^\prime\int d\fr\,\left(\dot{\Phi}_\gamma(\fr,t^\prime)-F_\gamma[\{\Phi_\mu(\fr,t^\prime)\}]\right)\\
&\qquad\times\int d\fr^\prime\,K^{-1}(\fr-\fr^\prime)\left(\dot{\Phi}_\gamma(\fr^\prime,t^\prime)-F_\gamma[\{\Phi_\mu(\fr^\prime,t^\prime)\}]\right),
\end{split}\label{eq:DefGenActionFunctional}
\end{align}
where $K^{-1}(\fr-\fr^\prime)$ is the inverse of the noise correlation kernel $K(\fr-\fr^\prime)$ from \eqref{eq:GeneralFTLangevin}. The two integral kernels fulfill 
\begin{equation}
\int d\fr^{\prime\prime}\,K(\fr-\fr^{\prime\prime})K^{-1}(\fr^{\prime\prime}-\fr^\prime)=\delta^d(\fr-\fr^\prime).\label{eq:DefGenRelationNoiseKernels}
\end{equation}
Note, that we do not explicitly state an expression for the Jacobian ensuing from the variable transformation $\eta(\fr,t)\to\Phi(\fr,t)$ in \eqref{eq:DefGenActionFunctional}. This is justified as we will use the action functional to derive a general expression for the medium entropy where it turns out that the Jacobian does not contribute (s.f. \cite{SeifertReview2012,SeifertStochTherm2008}). Inserting \eqref{eq:DefGenProbDensity}, \eqref{eq:DefGenActionFunctional} into \eqref{eq:DefGenMediumEntropy} and noticing that only the time-antisymmetric part of the action functional \eqref{eq:DefGenActionFunctional} and its time-reversed counterpart survives, leads to (see also \cite{SeifertReview2012,SeifertStochTherm2008,Maes2008})
\begin{equation}
\Delta s_\text{m}=2\sum_\gamma\int_{t_0}^tdt^\prime\int d\fr\int d\fr^\prime\,\dot{\Phi}_\gamma(\fr,t^\prime)K^{-1}(\fr-\fr^\prime)F_\gamma[\{\Phi_\mu(\fr^\prime,t^\prime)\}].\label{eq:DefGenMediumEntropyExplicit}
\end{equation}
The stochastic entropy change $\Delta s$ for the same trajectory, is given by (see also \cite{SeifertStochTherm2008})
\begin{equation}
\Delta s\equiv-\ln p[\fPhi(\fr,\tau)]\Big|_{t_0}^t.\label{eq:DefGenStochEntropy}
\end{equation}
Thus, the total entropy production $\Delta s_\text{tot}$ reads
\begin{align}
\begin{split}
\Delta s_\text{tot}&=2\sum_\gamma\int_{t_0}^tdt^\prime\int d\fr\int d\fr^\prime\,\dot{\Phi}_\gamma(\fr,t^\prime)K^{-1}(\fr-\fr^\prime)F_\gamma[\{\Phi_\mu(\fr^\prime,t^\prime)\}]\\
&\qquad-\ln p[\fPhi(\fr,\tau)]\Big|_{t_0}^t.
\end{split}\label{eq:DefGenTotalEntropyProd}
\end{align}
With \eqref{eq:DefGenTotalEntropyProd} we may also define the rate of total entropy production $\sigma$ in a NESS according to
\begin{equation}
\sigma=\lim_{t\to\infty}\frac{\expval{\Delta s_\text{tot}}}{t}.\label{eq:DefGenSigma}
\end{equation}
The expressions stated in \eqref{eq:DefGenEpsilonSquared} and \eqref{eq:DefGenTotalEntropyProd} provide us with the necessary ingredients to formulate the field-theoretic thermodynamic uncertainty relation as
\begin{equation}
\expval{\Delta s_\text{tot}}\,\epsilon^2=\frac{2\,D_g\,\sigma}{J_g^2}\geq2,\label{eq:DefGenFTTUR}
\end{equation}
with $\sigma$ from \eqref{eq:DefGenSigma}, $D_g$ from \eqref{eq:DefGenDiffusivity} and $J_g$ from \eqref{eq:DefGenProjectedCurrentSteadyState}. The higher the precision, i.e. the smaller $\epsilon^2$, the more entropy $\expval{\Delta s_\text{tot}}$ is generated, i.e. the higher the thermodynamic cost. Or, in other words, in order to sustain a certain NESS current $J_g$, a minimal entropy production rate $\sigma\geq J_g^2/D_g$ is required.

\section{Theoretical Background}\label{sec:TheoBack}

Within this section we will lay the groundwork for the calculation of the quantities entering the TUR for the KPZ equation. The main focus thereby is on the perturbative solution of the KPZ equation in the weak-coupling regime and the discussion of issues with diverging terms due to a lack of regularity.

\subsection{The KPZ Equation in Spectral Form}\label{subsec:KPZinSpectralForm}

Consider the one-dimensional KPZ equation \cite{KardarParisiZhang1986} on the interval $[0,b]$, $b>0$, with Gaussian white noise $\eta(x,t)$
\begin{align}
\begin{split}
\parderiv{h(x,t)}{t}&=\hat{L}\,h(x,t)+\frac{\lambda}{2}\,\left(\parderiv{h(x,t)}{x}\right)^2+\eta(x,t)\\
\expval{\eta(x,t)}&=0\\
\expval{\eta(x,t)\,\eta(x^\prime,t^\prime)}&=\Delta_0\delta(x-x^\prime)\delta(t-t^\prime),
\end{split}\label{eq:DefinitionKPZ}
\end{align}
subject to periodic boundary conditions and, for simplicity, vanishing initial condition $h(x,0)=0$, $x\in[0,b]$ (i.e. the growth process starts with a flat profile). Here $\hat{L}=\nu\partial_x^2$ is a differential diffusion operator, $\Delta_0$ a constant noise strength, and $\lambda$ the coupling constant of the non-linearity.\\
A Fourier-expansion of the height field $h(x,t)$ and the stochastic driving force $\eta(x,t)$ reads
\begin{align}
\begin{split}
h(x,t)&=\sum_{\inZ{k}}h_k(t)\phi_k(x),\\
\eta(x,t)&=\sum_{\inZ{k}}\eta_k(t)\phi_k(x).
\end{split}\label{eq:GenEigFuncExpansion}
\end{align}
The set of $\{\phi_k(x)\}$ is given by
\begin{equation}
\phi_k(x)\equiv \frac{1}{\sqrt{b}}e^{2\pi ikx/b}\qquad\inZ{k},\label{eq:DefUnscaledEigenFunc}
\end{equation}
and thus $h_k(t)$, $\eta_k(t)\in\mathds{C}$ in \eqref{eq:GenEigFuncExpansion}. A similar proceeding for the case of the Edwards-Wilkinson equation was used in \cite{ChouPleimlingZia2009,ChouPleimling2010,ChouPleimling2012,HenkelNohPleimling2012}. Inserting \eqref{eq:GenEigFuncExpansion} into \eqref{eq:DefinitionKPZ} leads to
\begin{align*}
&\sum_{\inZ{k}}\dot{h}_k(t)\phi_k(x)\\
&=\sum_{\inZ{k}}h_k(t)\hat{L}\phi_k(x)+\frac{\lambda}{2}\sum_{\inZ{l,m}}h_l(t)h_m(t)\partial_x\phi_l(x)\partial_x\phi_m(x)+\sum_{\inZ{k}}\eta_k(t)\phi_k(x)\\
&=\sum_{\inZ{k}}h_k(t)\mu_k\phi_k(x)-2\pi^2\frac{\lambda}{b^2}\sum_{\inZ{l,m}}l\,m\,h_l(t)h_m(t)\phi_l(x)\phi_m(x)\\
&\qquad+\sum_{\inZ{k}}\eta_k(t)\phi_k(x),
\end{align*}
with $\{\mu_k\}$ defined as
\begin{equation}
\mu_k\equiv -4\,\pi^2\,\frac{\nu}{b^2}\,k^2\qquad\inZ{k}.\label{eq:UnscaledEigenvaluesOfL}
\end{equation}
For the $\{\phi_k(x)\}$ the relation $\phi_l(x)\phi_m(x)=\phi_{l+m}(x)/\sqrt{b}$ holds and thus the double-sum in the Fourier expansion of the KPZ equation can be rewritten in convolution form setting $k=l+m$. This yields
\begin{align}
\begin{split}
&\sum_{\inZ{k}}\dot{h}_k(t)\phi_k(x)\\
&=\sum_{\inZ{k}}h_k(t)\mu_k\phi_k(x)-2\pi^2\frac{\lambda}{b^{5/2}}\sum_{\inZ{k,l}}l(k-l)h_l(t)h_{k-l}(t)\phi_k(x)\\
&\qquad+\sum_{\inZ{k}}\eta_k(t)\phi_k(x),
\end{split}\label{eq:SpectralFormKPZEquation}
\end{align}
which implies ordinary differential equations for the Fourier-coefficients $h_k(t)$,
\begin{equation}
\dot{h}_k(t)=\mu_kh_k(t)-2\pi^2\frac{\lambda}{b^{5/2}}\sum_{\inZ{l}}l(k-l)h_l(t)h_{k-l}(t)+\eta_k(t).\label{eq:DEQforHks}
\end{equation}
The above ODEs \eqref{eq:DEQforHks} are readily \lq solved\rq{} by the variation of constants formula, which leads for flat initial condition $h_k(0)\equiv0$ to
\begin{equation}
h_k(t)= \int_0^tdt^\prime e^{\mu_k(t-t^\prime)}\left[\eta_k(t^\prime)-2\pi^2\frac{\lambda}{b^{5/2}}\sum_{\inZwo{l}}l(k-l)h_l(t^\prime)h_{k-l}(t^\prime)\right],\label{eq:FlatMildSolOfKthFourierCoeff}
\end{equation}
$\inZ{k}$. Note, that the assumption of flat initial conditions is not in conflict with \eqref{eq:DefGenStochEntropy} as in the NESS, in which the relevant quantities will be evaluated, the probability density becomes stationary. With \eqref{eq:FlatMildSolOfKthFourierCoeff}, a non-linear integral equation for the $k$-th Fourier coefficient has been derived. In \autoref{subsec:SmallLambdaExpansion}, the solution to \eqref{eq:FlatMildSolOfKthFourierCoeff} will be constructed by means of an expansion in a small coupling parameter $\lambda$. We close this section with the following general remarks.\\
(i) Equation \eqref{eq:FlatMildSolOfKthFourierCoeff} has been derived on a purely formal level. In particular, the integral $\int dt^\prime e^{\mu_k(t-t^\prime)}\eta_k(t^\prime)$ has to be given a meaning. In a strict mathematical formulation, this integral has to be written as
\begin{equation}
\int_0^te^{\mu_k(t-t^\prime)}dW_k(t^\prime),\label{eq:DefStochConvolution}
\end{equation}
which is called a stochastic convolution (see e.g. \cite{DaPratoZabczykBook1992,EvansBook2002,DaPrato1994,DaPratoZabczykBook1996}). This has its origin in the fact that the noise $\eta(x,t)$ in \eqref{eq:DefinitionKPZ} is mathematically speaking a generalized time-derivative of a Wiener process $W(x,t)$ (see also \autoref{subsec:CloserLookAtNoise}, \eqref{eq:DefGenEta}). In this spirit, \eqref{eq:FlatMildSolOfKthFourierCoeff} with the first integral on the right hand side replaced by \eqref{eq:DefStochConvolution} may be called the mild form of the KPZ equation (in its spectral representation) and $h(x,t)=\sum_{\inZ{k}}h_k(t)\phi_k(x)$, $h_k(t)$ solution of equation \eqref{eq:FlatMildSolOfKthFourierCoeff}, is then called a mild solution of the KPZ equation. In mathematical literature, proofs of existence and uniqueness of such a mild solution can be found for various assumptions on the regularity of the noise (see e.g. \cite{DaPrato1994,Goldys2005,Bloemker2013} and references therein). An assumption will be adopted (see \autoref{subsec:CloserLookAtNoise}), which guarantees the existence of $\norm{h(x,t)}{([0,b])}$, i.e. the norm on the Hilbert space of square-integrable functions $\mathcal{L}_2$. This norm, or respectively the corresponding $\mathcal{L}_2$-product, denoted in the following by $(\cdot,\cdot)_0$, of $h$ with any $\mathcal{L}_2$-function $g$, i.e. $(h,g)_0$, will be used in \autoref{subsec:VarAndMeanHeightField} and \autoref{subsec:TotEntropyProductionKPZ} to calculate the necessary contributions to a field-theoretic thermodynamic uncertainty relation. Furthermore, with this assumption on the noise, it is shown in \autoref{app:MildSolution} for the mild solution that almost surely $h(x,t)\in\mathcal{C}([0,T],\mathcal{L}_2([0,b]))$, $T>0$, i.e. the trajectory $t\mapsto h(x,t)$ is a continuous function in time $t$ with values $h(\cdot,t)\in\mathcal{L}_2([0,b])$. This justifies the choice $H=\mathcal{L}_2([0,b])$ in the following calculations.\\
(ii) The Fourier expansion applied above can be understood in a more general sense. For the case of periodic boundary conditions, the differential operator $\hat{L}$ possesses the eigenfunctions $\{\phi_k(x)\}$ and corresponding eigenvalues $\{\mu_k\}$ from \eqref{eq:DefUnscaledEigenFunc} and \eqref{eq:UnscaledEigenvaluesOfL}, respectively. It is well-known that the set $\{\phi_k(x)\}$ constitutes a complete orthonormal system in the Hilbert space $\mathcal{L}_2(0,b)$ of all square-integrable functions on $(0,b)$. Thus the Fourier-expansion performed above can also be interpreted as an expansion in the eigenfunctions of the operator $\hat{L}$.\\
(iii) With this interpretation, \eqref{eq:FlatMildSolOfKthFourierCoeff} also holds for a \lq hyperdiffusive\rq{} version of the KPZ equation in which the operator $\hat{L}$ is replaced by $\hat{L}_p\equiv(-1)^{p+1}\partial_x^{2p}$, with $\inN{p}$ and adjusted eigenvalues $\{\mu_k^p\}$. This may be used to introduce a higher regularity to the KPZ equation.\\
(iv) Besides the complex Fourier expansion in \eqref{eq:GenEigFuncExpansion} with coefficients $h_k(t)\in\mathds{C}$, the real expansion $h(x,t)=\sum_{\inZ{k}}\wt{h}_k(t)\gamma_k(x)$, $\wt{h}_k(t)\in\mathds{R}$ (e.g. \cite{DaPrato1994}) and
\begin{equation}
\gamma_0=\frac{1}{\sqrt{b}},\quad\gamma_k=\sqrt{\frac{2}{b}}\sin2\pi k\frac{x}{b},\quad\gamma_{-k}=\sqrt{\frac{2}{b}}\cos2\pi k\frac{x}{b}\quad\inN{k},\label{eq:CosSinEigFun}
\end{equation} 
will be used in the next section. The relationship between $h_k(t)$ and $\wt{h}_k(t)$ reads
\begin{equation}
h_k(t)=\frac{\wt{h}_{-k}(t)-i\wt{h}_k(t)}{\sqrt{2}}\,,\qquad h_{-k}(t)=\frac{\wt{h}_{-k}(t)+i\wt{h}_k(t)}{\sqrt{2}}=\cc{h_k}(t),\label{eq:DefComplexFourierCoeff}
\end{equation}
with $\cc{h_k}(t)$ as the complex conjugate.

\subsection{A Closer Look at the Noise}\label{subsec:CloserLookAtNoise}

In the following discussion of the noise it is instructive to pretend, for the time being, that the noise is spatially colored with noise correlator $K(x-x^\prime)$ instead of assuming directly spatially white noise.\\
The noise $\eta(x,t)$ is given by a generalized time-derivative of a Wiener process $W(x,t)\in\mathds{R}$ \cite{DaPratoZabczykBook1992,EvansBook2002,DaPrato1994,FoiasBookNavierStokes2001}, i.e.
\begin{equation}
\eta(x,t)=\sqrt{\Delta_0}\,\parderiv{W(x,t)}{t}.\label{eq:DefGenEta}
\end{equation}
Such a Wiener process $W(x,t)$ can be written as (e.g. \cite{DaPratoZabczykBook1992,DaPrato1994})
\begin{equation}
W(x,t)=\sum_{\inZ{k}}\alpha_k\beta_k(t)\gamma_k(x).\label{eq:DefGenQWiener}
\end{equation}
Here $\{\alpha_k\}\in\mathds{R}$ are arbitrary expansion coefficients that may be used to introduce a spatial regularization of the Wiener process, $\{\beta_k(t)\}\in\mathds{R}$ are stochastically independent standard Brownian motions and $\{\gamma_k(x)\}$ from \eqref{eq:CosSinEigFun}. A well-known result for the two-point correlation function of two stochastically independent Brownian motions $\beta_k(t)$ reads \cite{DaPratoZabczykBook1992}
\begin{equation}
\expval{\beta_k(t)\,\beta_l(t^\prime)}=\delta_{k,l}\,(t\wedge t^\prime),\label{eq:GenExpValStandBrownianMotion}
\end{equation}
with $(t\wedge t^\prime)=\min(t,t^\prime)$.\\
In the following it will be shown that the noise $\eta$ defined by \eqref{eq:DefGenEta} and \eqref{eq:DefGenQWiener} possesses the autocorrelation
\begin{equation}
\expval{\eta(x,t)\,\eta(x^\prime,t^\prime)}=K(x-x^\prime)\delta(t-t^\prime),\label{eq:NoiseCorrelationRealSpace}
\end{equation}
which for $K(x-x^\prime)=\Delta_0\delta(x-x^\prime)$ results in the one assumed in \eqref{eq:DefinitionKPZ}. Furthermore, an explicit expression of the kernel $K(x-x^\prime)$ by means of the Fourier coefficients $\{\alpha_k\}$ of $W(x,t)$ from \eqref{eq:DefGenQWiener} will be given.\\
To this end, first an expression for the two-point correlation function of the Wiener process itself can be derived according to
\begin{align}
\begin{split}
\expval{W(x,t)\,W(x^\prime,t^\prime)}&=\frac{t\wedge t^\prime}{b}\left[\alpha_0^2+\sum_{\inN{k}}\left[\alpha_{-k}^2+\alpha_k^2\right]\cos 2\pi k\frac{x-x^\prime}{b}\right.\\
&\qquad\left.+\sum_{\inN{k}}\left[\alpha_{-k}^2-\alpha_k^2\right]\cos 2\pi k\frac{x+x^\prime}{b}\right].\label{eq:GenDefTwoPointCorrWienerProcess}
\end{split}
\end{align}
To represent the noise structure dictated by \eqref{eq:DefinitionKPZ}, the expression in \eqref{eq:GenDefTwoPointCorrWienerProcess} has to be an even, translationally invariant function in space. Thus, the following relation has to be fulfilled
\begin{equation}
\alpha_{-k}=\alpha_k\quad\forall\,\inN{k}.\label{eq:RelationAlphaKs}
\end{equation}
Then the two-point correlation function of the Wiener process is given by
\begin{equation}
\expval{W(x,t)\,W(x^\prime,t^\prime)}=\frac{t\wedge t^\prime}{b}\left[\alpha_0^2+2\sum_{\inN{k}}\alpha_k^2\,\cos 2\pi k\frac{x-x^\prime}{b}\right].\label{eq:TwoPointCorrWienerProcess}
\end{equation}
With $W(x,t)=\sum_{\inZ{k}}W_k(t)\phi_k(x)$, $\phi_k(x)$ from \eqref{eq:DefUnscaledEigenFunc}, equation \eqref{eq:TwoPointCorrWienerProcess} implies for the two-point correlation function of the Fourier coefficients $W_k(t)$
\begin{equation}
\expval{W_k(t)\,W_l(t^\prime)}=\alpha_k\,\alpha_l\,\delta_{k,-l}\,(t\wedge t^\prime),\qquad \inZ{k,l}.\label{eq:CorrFuncWks}
\end{equation}
This result leads immediately to 
\begin{equation}
\expval{\eta_k(t)\,\eta_l(t^\prime)}\equiv\Delta_0\parderivTwo{\expval{W_k(t)\,W_l(t^\prime)}}{t}{t^\prime}=\Delta_0\alpha_k\,\alpha_l\,\delta_{k,-l}\,\delta(t-t^\prime),\qquad \inZ{k,l},\label{eq:CorrFuncEtaKs}
\end{equation}
using $\partial_t\partial_{t^\prime}(t\wedge t^\prime)=\delta(t-t^\prime)$.\\
For the relation between \eqref{eq:TwoPointCorrWienerProcess} and the noise from \eqref{eq:NoiseCorrelationRealSpace}, we differentiate \eqref{eq:TwoPointCorrWienerProcess} with respect to $t$ and $t^\prime$ yielding
\begin{align}
\begin{split}
\expval{\eta(x,t)\,\eta(x^\prime,t^\prime)}&=\Delta_0\parderivTwo{\expval{W(x,t)\,W(x^\prime,t^\prime)}}{t}{t^\prime}\\
&=\frac{\Delta_0}{b}\left[\alpha_0^2+2\sum_{\inN{k}}\alpha_k^2\,\cos 2\pi k\frac{x-x^\prime}{b}\right]\delta(t-t^\prime).\label{eq:DefTwoPointCorrEta}
\end{split}
\end{align}
The following identification can be made
\begin{equation}
K(x-x^\prime)=\frac{\Delta_0}{b}\left[\alpha_0^2+2\sum_{\inN{k}}\alpha_k^2\,\cos 2\pi k\frac{x-x^\prime}{b}\right]=K(|x-x^\prime|),\label{eq:DefFourierExpNoiseKernel}
\end{equation}
which structurally represents the standard implicit assumption that $K(x-x^\prime)$ is translationally invariant, positive definite and even. Note, that the regularity of the noise-kernel $K(|x-x^\prime|)$ is given by the behavior of the set of $\{\alpha_k\}$ for $k\to\infty$, where $\{\alpha_k\}$ are the dimensionless Fourier coefficients of the underlying Wiener process from \eqref{eq:DefGenQWiener} for all $k$. For the case of $\alpha_k=1$ $\forall\;\inZ{k}$, spatially white noise is obtained.\\
Thus, the derivation via the Wiener process has indeed led to a translationally invariant real-valued two-point correlation function for $\eta(x,t)$, given by \eqref{eq:NoiseCorrelationRealSpace}, with $K(|x-x^\prime|)$ from \eqref{eq:DefFourierExpNoiseKernel}, which describes white in time and spatially colored Gaussian noise. In the following, we will use \eqref{eq:DefFourierExpNoiseKernel} to approximate spatially white noise to meet the required form in \eqref{eq:DefinitionKPZ}.\\
Now the assumption mentioned in the remarks in \autoref{subsec:KPZinSpectralForm} can be made more precise. In the following it will be assumed that (see \autoref{app:MildSolution})
\begin{equation}
\sum_{\inZ{k}}k^\epsilon\alpha_k^2<\infty,\qquad\epsilon>0.\label{eq:ConvergenceCondAlphaKs}
\end{equation}
This assumption excludes white noise for $\inN{k}$, but via the introduction of a cutoff parameter $\inN{\Lambda}$, $\Lambda\gg1$ arbitrarily large but finite, for the range of $k$, white noise is accessible, i.e. for $\inR{k}$ with 
\begin{equation}
\mathfrak{R}\equiv[-\Lambda,\Lambda].\label{eq:DefR}
\end{equation}
Note that for the linear case, i.e. the Edwards-Wilkinson model, the authors of \cite{ChouPleimling2010} also introduce a cutoff, albeit in a slightly different manner. Such a cutoff amounts to an orthogonal projection of the full eigenfunction expansion of \eqref{eq:DefinitionKPZ} to a finite-dimensional subspace spanned by the eigenfunctions $\phi_{-\Lambda}(x),\ldots,\phi_\Lambda(x)$. Mathematically, this projection may be represented by a linear projection operator $\mathcal{P}_\Lambda$, which maps the Hilbert space $\mathcal{L}_2(0,b)$ to $\text{span}\{\phi_{-\Lambda}(x),\ldots,\phi_\Lambda(x)\}$, acting on \eqref{eq:SpectralFormKPZEquation}. This mapping, however, causes a problem in the non-linear term of \eqref{eq:SpectralFormKPZEquation}, where by mode coupling the $k$-th Fourier mode ($-\Lambda\leq k\leq \Lambda$) is influenced also by modes with $|l|>\Lambda$. This issue can be resolved by choosing $\Lambda$ large enough, for modes with $h_l(t)\sim\exp[\mu_lt]$, $\mu_l$ from \eqref{eq:UnscaledEigenvaluesOfL}, \eqref{eq:EigenvaluesOfL}, $|l|>\Lambda$ will be damped out rapidly so that the bias introduced by limiting $l$ to the interval $\mathfrak{R}$ is small. Note that the restriction to $h\in\text{span}\{\phi_{-\Lambda},\,\ldots,\,\phi_\Lambda\}$ also implies the introduction of restricted summation boundaries in the convolution term in \eqref{eq:FlatMildSolOfKthFourierCoeff}, namely
\begin{align*}
\sum_{\inZ{l}}l(k-l)h_lh_{k-l}\quad\longrightarrow\quad\sum_{\inRkWO{l}{k}{,k}}l(k-l)h_lh_{k-l},\qquad\inR{k},
\end{align*}
with $\mathfrak{R}_k$ defined by
\begin{equation}
\mathfrak{R}_k\equiv[\max(-\Lambda,-\Lambda+k),\min(\Lambda,\Lambda+k)]\,,\qquad \inR{k}.\label{eq:DefRk}
\end{equation}
This restriction to finitely many Fourier modes is not as harsh as it might seem, since for very large wavenumbers the dynamics of the KPZ equation is governed by the Edwards-Wilkinson equation, which, due to its equilibrium behavior, does not contribute to the thermodynamic uncertainty relation \eqref{eq:DefGenFTTUR} (see e.g. \cite{FogedbyBurgers1998,Fogedby2002,FogedbyKPZ2006,Fogedby2008}).\\
With the cutoff $\Lambda$, condition \eqref{eq:ConvergenceCondAlphaKs} is of course fulfilled for $\alpha_k=1$ $\forall\,\inR{k}$ and $\alpha_k=0$ $\forall k$ $\notin\mathfrak{R}$. Inserting this choice of $\alpha_k$ into \eqref{eq:DefFourierExpNoiseKernel} yields
\begin{equation}
K(x-x^\prime)=\frac{\Delta_0}{b}\left[1+2\sum_{k=1}^\Lambda\cos 2\pi k\frac{x-x^\prime}{b}\right]=\Delta_0\delta(x-x^\prime)\Big|_{\text{span}\{\phi_{-\Lambda},\ldots,\phi_\Lambda\}}.\label{eq:DefFourierExpNoiseKernelWhite}
\end{equation}
Also, the choice of $\alpha_k=1$ $\forall\,\inR{k}$ implies for the correlation function of the Fourier coefficients $\eta_k(t)$ from \eqref{eq:CorrFuncEtaKs}
\begin{equation}
\expval{\eta_k(t)\eta_l(t^\prime)}=\Delta_0\delta_{k,-l}\delta(t-t^\prime)\qquad\inR{k,l}.\label{eq:CorrFuncEtaKsWhite}
\end{equation}
To end this section, a noise operator $\hat{K}$ describing spatial noise correlations will be introduced as
\begin{equation}
\hat{K}(\cdot)\equiv\int_0^bdx^\prime K(x-x^\prime)(\cdot)(x^\prime),\label{eq:DefNoiseOperator}
\end{equation}
with kernel $K(x-x^\prime)$ from \eqref{eq:DefFourierExpNoiseKernelWhite} and its inverse $\hat{K}^{-1}$ given by
\begin{equation}
\hat{K}^{-1}(\cdot)=\int_0^bdx^\prime K^{-1}(x-x^\prime)(\cdot)(x^\prime),\label{eq:DefInvNoiseOperator}
\end{equation}
where its kernel reads $K^{-1}(x-x^\prime)=\Delta_0^{-1}\delta(x-x^\prime)\Big|_{\text{span}\{\phi_{-\Lambda},\ldots,\phi_\Lambda\}}$.

\subsection{Dimensionless Form of the KPZ Equation}\label{subsec:DimlessFormKPZ}

Before the KPZ equation is analyzed further, it is prudent to relate all physical quantities to suitable reference values so that the scaled quantities are dimensionless and that the equation is characterized by only one dimensionless parameter. In anticipation of the calculations below, we choose this parameter to represent a dimensionless effective coupling parameter $\lambda_\text{eff}$, that replaces the coupling constant $\lambda$ from \eqref{eq:DefinitionKPZ}. To this end the following characteristic scales are introduced,
\begin{equation}
h=Hh_\text{s}\;;\quad\eta=N\eta_\text{s}\;;\quad x=b x_\text{s}\;;\quad t=Tt_\text{s}.\label{eq:ScalingRelations}
\end{equation}
Here $H$ is a characteristic scale for the height field (not to be confused with the notation for the Hilbert space), $N$ a scale for the noise field, $b$ is the characteristic length scale in space and $T$ the time scale of the system.
Choosing the three respective scales according to
\begin{equation}
H=\sqrt{\frac{\Delta_0\,b}{\nu}},\quad N=\sqrt{\frac{\Delta_0\,\nu}{b^3}},\quad T=\frac{b^2}{\nu},\label{eq:DefScales}
\end{equation}
leads to the dimensionless KPZ equation on the interval $x\in[0,1]$
\begin{align}
\partial_{t_\text{s}}h_\text{s}(x_\text{s},t_\text{s})&=\partial_{x_\text{s}}^2h_\text{s}(x_\text{s},t_\text{s})+\frac{\lambda_\text{eff}}{2}\left(\partial_{x_\text{s}}h_\text{s}(x_\text{s},t_\text{s})\right)^2+\eta_\text{s}(x_\text{s},t_\text{s}),\label{eq:RescaledKPZDimensionless}\\
\expval{\eta_\text{s}(x_\text{s},t_\text{s})}&=0,\label{eq:RescaledExpValEta}\\
\expval{\eta_\text{s}(x_\text{s},t_\text{s})\eta_\text{s}(x_\text{s}^\prime,t_\text{s}^\prime)}&=K_\text{s}(x_\text{s}-x_\text{s}^\prime)\delta(t_\text{s}-t_\text{s}^\prime).\label{eq:RescaledAutoCorrEta}
\end{align}
Here, the effective dimensionless coupling constant is given by
\begin{equation}
\lambda_\text{eff}=\frac{\lambda\,\Delta_0^{1/2}}{\nu^{3/2}}b^{1/2},\label{eq:DefEffCouplingConst}
\end{equation}
and
\begin{equation}
K_\text{s}(x_\text{s}-x_\text{s}^\prime)=1+2\sum_{k=1}^\Lambda\cos 2\pi k(x_\text{s}-x^\prime_\text{s}) \label{eq:RescaledFourierExpNoiseKernel}. 
\end{equation}
The effective coupling constant $\lambda_\text{eff}$ is found in various works concerning the KPZ-Burgers equation; see e.g. \cite{ForsterNelsonStephen1977,FreyTaeuber1994,MedinaHwaKardarZhang1989,NiggemannSincNoise2018}.\par
In the following sections we will perform all calculations for the dimensionless KPZ equation. This requires one simple adjustment in the linear differential operator $\hat{L}$ on $x_\text{s}\in[0,1]$, which is now given by
\begin{equation}
\hat{L}_\text{s}=\partial_{x_\text{s}}^2,\label{eq:DefDimLessOperatorL}
\end{equation}
with eigenvalues 
\begin{equation}
\mu_{s,\,k}=-4\,\pi^2\,k^2\label{eq:EigenvaluesOfL}
\end{equation}
to the orthonormal eigenfunctions
\begin{equation}
\phi_{s,\,k}(x_\text{s})=e^{2\pi ikx_\text{s}}.\label{eq:DefEigenFunc}
\end{equation}
Furthermore, the noise correlation function in Fourier space from \eqref{eq:CorrFuncEtaKsWhite} now reads
\begin{equation}
\expval{\eta_{s,\,k}(t_\text{s})\eta_{s,\,l}(t_\text{s}^\prime)}=\delta_{k,-l}\delta(t_\text{s}-t_\text{s}^\prime).\label{eq:RescaledCorrFuncEtaKs}
\end{equation}
The scaling also effects the noise operators defined in \eqref{eq:DefNoiseOperator}, \eqref{eq:DefInvNoiseOperator} at the end of \autoref{subsec:CloserLookAtNoise}. The scaled ones read
\begin{equation}
\hat{K}_\text{s}(\cdot)=\int_0^1dx_\text{s}^\prime\,K_\text{s}(x_\text{s}-x^\prime_\text{s})(\cdot)(x^\prime_\text{s}),\label{eq:DefNoiseOperatorScaled}
\end{equation}
and 
\begin{equation}
\hat{K}^{-1}_\text{s}(\cdot)=\int_0^1dx_\text{s}^\prime\,K_\text{s}^{-1}(x_\text{s}-x_\text{s}^\prime)(\cdot)(x_\text{s}^\prime),\label{eq:DefInvNoiseOperatorScaled}
\end{equation}
with $K_\text{s}(x_\text{s}-x_\text{s}^\prime)$ from \eqref{eq:RescaledFourierExpNoiseKernel} and $K^{-1}_\text{s}(x_\text{s}-x^\prime_\text{s})$ is defined via the integral-relation $\int dy_\text{s}\,K_\text{s}(x_\text{s}-y_\text{s})K_\text{s}^{-1}(y_\text{s}-z_\text{s})=\delta(x_\text{s}-z_\text{s})$.\\
Note that for the sake of simplicity the subscript $\text{s}$ will be dropped in the calculations below where all quantities are understood as the scaled ones.

\subsection{Expansion in a Small Coupling Constant}\label{subsec:SmallLambdaExpansion}

Returning to the nonlinear integral equation of the $k$-th Fourier coefficient of the heights field, $h_k(t)$ from \eqref{eq:FlatMildSolOfKthFourierCoeff}, now in its dimensionless form and with the restricted spectral range given by
\begin{equation}
h_k(t)= \int_0^tdt^\prime e^{\mu_k(t-t^\prime)}\left[\eta_k(t^\prime)-2\pi^2\lambda_\text{eff}\sum_{\inRkWO{l}{k}{,k}}l(k-l)h_l(t^\prime)h_{k-l}(t^\prime)\right],\label{eq:FlatMildSolOfKthFourierCoeffScaled}
\end{equation}
$\inR{k}$, with $\{\mu_k\}$ from \eqref{eq:EigenvaluesOfL}, $\mathfrak{R}_k$ from \eqref{eq:DefRk} and all quantities dimensionless, an approximate solution will be constructed. Note, that the summation of the discrete convolution in \eqref{eq:FlatMildSolOfKthFourierCoeffScaled} is chosen such that it respects the above introduced cutoff in $l$ as well as $k-l$,  i.e. $|l|$, $|k-l|\leq\Lambda$. For small values of the coupling constant we expand the solution in powers of $\lambda_\text{eff}$, i.e.
\begin{equation}
h_k(t)=h_k^{(0)}(t)+\lambda_\text{eff} h_k^{(1)}(t)+\lambda_\text{eff}^2h_k^{(2)}(t)+O(\lambda_\text{eff}^3),\label{eq:GenDefSmallLambdaExpansion}
\end{equation}
with
\begin{align}
\hh{k}{0}(t)&=\int_0^t e^{\mu_k(t-t^\prime)}dW_k(t^\prime),\label{eq:DefHk0}\\
\hh{k}{1}(t)&=-2\pi^2\sum_{\inRkWO{l}{k}{,k}}l(k-l)\int_0^tdt^\prime e^{\mu_k(t-t^\prime)}\hh{l}{0}(t^\prime)\hh{k-l}{0}(t^\prime),\label{eq:DefHk1}\\
\begin{split}
\hh{k}{2}(t)&=-2\pi^2\sum_{\inRkWO{l}{k}{,k}}l(k-l)\int_0^tdt^\prime e^{\mu_k(t-t^\prime)}\\
&\qquad\times\left(\hh{l}{0}(t^\prime)\hh{k-l}{1}(t^\prime)+\hh{l}{1}(t^\prime)\hh{k-l}{0}(t^\prime)\right)
\end{split}\label{eq:DefHk2}
\end{align}
Thus every $\hh{k}{n}$, $n>1$, can be expressed in terms of $\hh{m}{0}$, $\inR{m}$, i.e. the stochastic convolution according to \eqref{eq:DefStochConvolution}, which is known to be Gaussian.\\
In the following calculations multipoint correlation functions have to be evaluated, which can be simplified by Wick's theorem, where a recurring term reads $\expval{\hh{k}{0}(t)\hh{l}{0}(t^\prime)}$. It is thus helpful to determine this correlation function in general once and use this result later on. With \eqref{eq:RescaledCorrFuncEtaKs} and $\inZ{k,l}$ (and therefore also for $\inR{k,l}$) it follows that:
\begin{align*}
\expval{\hh{k}{0}(t)\hh{l}{0}(t^\prime)}&=e^{\mu_kt}e^{\mu_lt^\prime}\int_0^tdr\int_0^{t^\prime}ds\,e^{-\mu_kr}e^{-\mu_ls}\expval{\eta_k(r)\eta_l(s)}\\
&=e^{\mu_kt}e^{\mu_lt^\prime}\delta_{k,-l}\frac{1-e^{-(\mu_k+\mu_l)(t\wedge t^\prime)}}{\mu_k+\mu_l}=\Pi_{k,l}(t,t^\prime)\delta_{k,-l},
\end{align*}
with 
\begin{equation}
\Pi_{k,l}(t,t^\prime)\equiv e^{\mu_kt}e^{\mu_lt^\prime}\frac{1-e^{-(\mu_k+\mu_l)(t\wedge t^\prime)}}{\mu_k+\mu_l}.\label{eq:DefAuxPsis}
\end{equation}
Since for the auxiliary expression $\Pi_{k,l}$ the symmetries 
\begin{equation}
\Pi_{k,l}(t,t^\prime)=\Pi_{k,-l}(t,t^\prime)=\Pi_{-k,l}(t,t^\prime)=\Pi_{-k,-l}(t,t^\prime)\label{eq:SymmetryPsis}
\end{equation}
hold, it is found that
\begin{align}
\begin{split}
\expval{\hh{k}{0}(t)\hh{l}{0}(t^\prime)}&=\expval{\cc{\hh{k}{0}}(t)\cc{\hh{l}{0}}(t^\prime)}=\Pi_{k,l}(t,t^\prime)\delta_{k,-l};\\
\expval{\hh{k}{0}(t)\cc{\hh{l}{0}}(t^\prime)}&=\expval{\cc{\hh{k}{0}}(t)\hh{l}{0}(t^\prime)}=\Pi_{k,l}(t,t^\prime)\delta_{k,l}.
\end{split}\label{eq:DefCorrFuncHks}
\end{align}

\section{Thermodynamic Uncertainty Relation for the KPZ Equation}\label{sec:ThermoUncertRel}

In this section we will show that the thermodynamic uncertainty relation from \eqref{eq:DefGenFTTUR} holds for the KPZ equation driven by Gaussian white noise in the weak-coupling regime. In particular, the small-$\lambda_\text{eff}$ expansion from \autoref{subsec:SmallLambdaExpansion} will be employed.\\
To recapitulate, the two ingredients needed for the thermodynamic uncertainty relation are (i) the long time behavior of the squared variation coefficient or precision $\epsilon^2$ of $\Psi_g(t)$ from \eqref{eq:DefGenEpsilonSquared}; (ii) the expectation value of the total entropy production in the steady state, $\expval{\Delta s_\text{tot}}$ from \eqref{eq:DefGenTotalEntropyProd}.

\subsection{Expectation and Variance for the Height Field}\label{subsec:VarAndMeanHeightField}

With \eqref{eq:DefGenProjectedOutputPsig} adapted to the KPZ equation, namely
\begin{equation}
\Psi_g(t)=\int_0^1dx\,h(x,t)g(x),\label{eq:DefGenProjectedOutputPsigKPZ}
\end{equation}
with $g(x)$ as any real-valued $\mathcal{L}_2$-function fulfilling $\int_0^1dxg(x)\neq0$, i.e. $g(x)$ possessing non-zero mean, we rewrite the variance as
\begin{equation}
\expval{\left(\Psi_g(t)-\expval{\Psi_g(t)}\right)^2}=\expval{\left(\Psi_g(t)\right)^2}-\expval{\Psi_g(t)}^2.\label{eq:DefVarGen}
\end{equation}
As is shown below, $\epsilon^2$ can be evaluated for arbitrary time $t>0$. However, the final interest is on the non-equilibrium steady state of the system. Therefore, the long-time asymptotics will be studied.

\paragraph{Evaluation of Expectation and Variance}\label{par:EvalMeanAndVar}

In the small $\lambda_\text{eff}$ expansion, the expectation of the output $\Psi_g(t)$ from \eqref{eq:DefGenProjectedOutputPsigKPZ}, with $h(x,t)$ solution of the dimensionless KPZ equation \eqref{eq:RescaledKPZDimensionless} to \eqref{eq:RescaledAutoCorrEta} reads:
\begin{align}
\begin{split}
\expval{\Psi_g(t)}&=\sum_{\inR{k,l}}\expval{h_k(t)}\cc{g_l}\left(e^{2\pi ikx},e^{2\pi ilx}\right)_0\\
&=\lambda_\text{eff}\sum_{\inR{k}}\cc{g_k}\expval{\hh{k}{1}(t)}+O(\lambda_\text{eff}^3),
\end{split}\label{eq:MeanSquaredGen}
\end{align}
where $g_k$ and $\cc{g_k}$ are the $k$-th Fourier coefficient of the weight function $g(x)$ and its complex conjugate, respectively. Here the result from \eqref{eq:DefHk0} is used as well as the fact that odd moments of Gaussian random variables vanish identically. Replacing $\hh{k}{1}(t)$ by the expression derived in \eqref{eq:DefHk1} and using \eqref{eq:DefCorrFuncHks} leads to
\begin{align}
\begin{split}
\expval{\hh{k}{1}(t)}&=-2\pi^2e^{\mu_kt}\int_0^tdt^\prime\,e^{-\mu_kt^\prime}\sum_{\inRkWO{l}{k}{,k}}l(k-l)\expval{\hh{l}{0}(t^\prime)\hh{k-l}{0}(t^\prime)}\\
&=-2\pi^2\sum_{\inRkWO{l}{k}{,k}}l(k-l)\left[\frac{e^{(\mu_l+\mu_{k-l})t}-e^{\mu_kt}}{(\mu_l+\mu_{k-l})(\mu_l+\mu_{k-l}-\mu_k)}\right.\\
&\qquad\left.-\frac{e^{\mu_kt}-1}{(\mu_l+\mu_{k-l})\mu_k}\right]\delta_{0,k}.\label{eq:ExpValHk1Interm1}
\end{split}
\end{align}
Note, that in the case of $k=0$ the second term in the last line of \eqref{eq:ExpValHk1Interm1} is evaluated in the limit $\mu_k\to0$, which yields $t$. Since the interest is on the steady state behavior, the long-time asymptotics of the two expressions in \eqref{eq:ExpValHk1Interm1} above is studied. So, eq. \eqref{eq:MeanSquaredGen} yields
\begin{equation}
\expval{\Psi_g(t)}\simeq 2\pi^2g_0\lambda_\text{eff}\sum_{\inRWO{l}{}}\frac{l^2}{2(-\mu_l)}\,t+O(\lambda_\text{eff}^3),\quad\text{for}\;t\gg1,\label{eq:MeanKPZ}
\end{equation}
where $g_k=g_{-k}$ $\forall k$ as $g(x)\in\mathds{R}$. Using the explicit form of $\mu_k$ from \eqref{eq:EigenvaluesOfL}, the expression in \eqref{eq:MeanKPZ} can be simplified according to
\begin{equation}
\expval{\Psi_g(t)}=g_0\frac{\lambda_\text{eff}}{2}\,\Lambda\,t+O(\lambda_\text{eff}^3),\quad\text{for}\;t\gg1,\label{eq:MeanKPZSimplified}
\end{equation}
with $\Lambda$ from \eqref{eq:DefR}. Equivalently, the steady state current from \eqref{eq:DefGenProjectedCurrentSteadyState} reads
\begin{equation}
J_g=g_0\frac{\lambda_\text{eff}}{2}\,\Lambda+O(\lambda_\text{eff}^3).\label{eq:StatCurrentKPZ}
\end{equation}
The first term of the variance as defined in \eqref{eq:DefVarGen} reads in the small-$\lambda_\text{eff}$ expansion
\begin{align}
\begin{split}
\expval{\left(\Psi_g(t)\right)^2}&=\expval{\sum_{\inR{k,l}}h_k(t)\cc{g_k}h_l(t)\cc{g_l}}\\
&=\sum_{\inR{k,l}}\cc{g_k}\,\cc{g_l}\left[\expval{\hh{k}{0}(t)\hh{l}{0}(t)}+\lambda_\text{eff}^2\left(\expval{\hh{k}{1}(t)\hh{l}{1}(t)}\right.\right.\\
&\qquad\left.\left.+\expval{\hh{k}{0}(t)\hh{l}{2}(t)}+\expval{\hh{k}{2}(t)\hh{l}{0}(t)}\right)+O(\lambda_\text{eff}^4)\right],
\end{split}\label{eq:ExpvalHkSquaredGen}
\end{align}
where moments proportional to $\lambda_\text{eff}$ (and $\lambda_\text{eff}^3$) vanish due to \eqref{eq:DefHk0} and \eqref{eq:DefHk1} as the two-point correlation function $\expval{\hh{k}{0}\hh{l}{1}}$ and its complex conjugate are odd moments.\\
In \autoref{app:EvaluationExpValHkSquared}, we present the rather technical derivation of
\begin{align}
\begin{split}
\expval{\left(\Psi_g(t)\right)^2}&\simeq g_0^2\left[1-2(2\pi^2)^2\lambda_\text{eff}^2\sum_{\inRWO{l}{}}\frac{l^4}{8\mu_l^3}\right]\,t\\
&\qquad+g_0^2\lambda_\text{eff}^2\sum_{\inR{k}}\left|\expval{\hh{k}{1}(t)}\right|^2+O(\lambda_\text{eff}^4)\quad\text{for}\;t\gg1.
\end{split}\label{eq:ExpvalHkSquaredKPZ}
\end{align}
Subtraction of \eqref{eq:MeanKPZ} squared from \eqref{eq:ExpvalHkSquaredKPZ} leads to
\begin{align}
\begin{split}
&\expval{\left(\Psi_g(t)\right)^2}-\expval{\Psi_g(t)}^2\\
&\simeq g_0^2\left[1-2(2\pi^2)^2\lambda_\text{eff}^2\sum_{\inRWO{l}{}}\frac{l^4}{8\mu_l^3}\right]\,t+O(\lambda_\text{eff}^4),\quad\text{for}\;t\gg1.
\end{split}\label{eq:VarianceKPZ}
\end{align}
Again, with $\mu_k$ from \eqref{eq:EigenvaluesOfL}, the above expression in \eqref{eq:VarianceKPZ} can be reduced to
\begin{equation}
\expval{\left(\Psi_g(t)\right)^2}-\expval{\Psi_g(t)}^2=g_0^2\left[1+\frac{\lambda_\text{eff}^2}{32\,\pi^2}\mathcal{H}_\Lambda^{(2)}\right]t+O(\lambda_\text{eff}^4).\label{eq:VarianceKPZSimplified}
\end{equation}
Here $\mathcal{H}_\Lambda^{(2)}=\sum_{l=1}^\Lambda1/l^2$ is the so-called generalized harmonic number, which converges to the Riemann zeta-function $\zeta(2)$ for $\Lambda\to\infty$. Using \eqref{eq:DefGenDiffusivity}, eq. \eqref{eq:VarianceKPZSimplified} yields the diffusivity $D_g$,
\begin{equation}
D_g=\frac{g_0^2}{2}\left[1+\frac{\lambda_\text{eff}^2}{32\,\pi^2}\mathcal{H}_\Lambda^{(2)}\right]+O(\lambda_\text{eff}^4).\label{eq:DiffusivityKPZ}
\end{equation}
With \eqref{eq:VarianceKPZSimplified} and \eqref{eq:MeanKPZSimplified} squared, the first constituent of the thermodynamic uncertainty relation, $\epsilon^2=\text{Var}[\Psi_g(t)]/\expval{\Psi_g(t)}^2$ from \eqref{eq:DefGenEpsilonSquared}, is given for large times by
\begin{equation}
\epsilon^2\simeq\frac{4+\lambda_\text{eff}^2/(8\pi^2)\mathcal{H}_\Lambda^{(2)}}{\lambda_\text{eff}^2\,\Lambda^2}\frac{1}{t}.\label{eq:EpsilonSquaredKPZ}
\end{equation}
Note, since $\epsilon^2\approx4/(\lambda_\text{eff}^2t)$, the long time asymptotics of the second term has to scale as $\expval{\Delta s_\text{tot}}\sim\lambda_\text{eff}^2t$ for the uncertainty relation to hold. Note further, that the result for the precision of the projected output $\Psi_g(t)$ in the NESS is independent of the choice of $g(x)$.

\subsection{Alternative Formulation of the Precision}\label{subsec:EpsilonSquaredNorm}

Before we continue with the calculation of the total entropy production, we would like to mention an intriguing observation. From the field-theoretic point of view, it seems natural to define the precision $\epsilon^2$ as
\begin{equation}
\epsilon^2\equiv\frac{\expval{\normZero{h(x,t)-\expval{h(x,t)}}^2}}{\normZero{\expval{h(x,t)}}^2}.\label{eq:EpsilonSquaredNorm}
\end{equation}
This is due to the fact that the height field $h(x,t)$ is at every time instance an element of the Hilbert-space $\mathcal{L}_2([0,1])$ as mentioned in \autoref{subsec:KPZinSpectralForm}. Hence, the difference between $h(x,t)$ and its expectation is measured by its $\mathcal{L}_2$-norm. Also the expectation squared is in this framework given by the $\mathcal{L}_2$-norm squared. At a cursory glance, the definitions in \eqref{eq:EpsilonSquaredNorm} and \eqref{eq:DefGenEpsilonSquared} seem to be incompatible. However, for the case of the above calculations of $\epsilon^2$ for the one-dimensional KPZ equation, it holds up to $O(\lambda_\text{eff}^3)$ in perturbation expansion that
\begin{align}
\begin{split}
\expval{\left(\Psi_g(t)-\expval{\Psi_g(t)}\right)^2}&=g_0^2\expval{\normZero{h(x,t)-\expval{h(x,t)}}^2}\qquad\text{for }t\gg1,\\
\expval{\Psi_g(t)}^2&=g_0^2\normZero{\expval{h(x,t)}}^2.
\end{split}\label{eq:EquivalenceProjectionNorm}
\end{align}
Thus, with \eqref{eq:EquivalenceProjectionNorm}, it is obvious that in terms of the perturbation expansion both definitions of the precision, as in \eqref{eq:DefGenEpsilonSquared} and \eqref{eq:EpsilonSquaredNorm}, respectively, are equivalent. Equation \eqref{eq:EquivalenceProjectionNorm} can be verified by direct calculation along the same lines as above in this section. By studying these calculations it is found perturbatively that the height field $h(x,t)$ is spatially homogeneous, which is reflected by $\expval{h_k(t)h_l(t)}\sim\delta_{k,-l}$ (see \eqref{eq:DefCorrFuncHks}) for the correlation of its Fourier-coefficients. Further, the long-time behavior is solely determined by the largest eigenvalue of the differential diffusion operator $\hat{L}=\partial_x^2$, namely by $\mu_0=0$ (see e.g. \eqref{eq:MeanKPZ} and \eqref{eq:VarianceKPZ}, the essential quantities for deriving \eqref{eq:EquivalenceProjectionNorm}).\\
In the following, we would like to give some reasoning why the above two statements should also hold for a broad class of field-theoretic Langevin equations as in \eqref{eq:GeneralFTLangevin}. For simplicity, we restrict ourselves in \eqref{eq:GeneralFTLangevin} to the case of one-dimensional scalar fields $\Phi(x,t)$ and $F[\Phi(x,t)]=\hat{L}\Phi(x,t)+\hat{N}[\Phi(x,t)]$. Here $\hat{L}$ denotes a linear differential operator and $\hat{N}$ a non-linear (e.g. quadratic) operator. $\hat{L}$ should be selfadjoint and possess a pure point spectrum with all eigenvalues $\mu_k\leq0$ (e.g. $\hat{L}=(-1)^{p+1}\partial_x^{2p}$, $\inN{p}$, i.e. an arbitrary diffusion operator subject to periodic boundary conditions). For this class of operators $\hat{L}$ there exists a complete orthonormal system of corresponding eigenfunctions $\{\phi_k\}$ in $\mathcal{L}_2(\Omega)$. If it is further known, that the solution $\Phi(x,t)$ of \eqref{eq:GeneralFTLangevin} belongs at every time $t$ to $\mathcal{L}_2(\Omega)$, we can calculate e.g. the second moment of the projected output $\Psi_g(t)$ according to $\expval{(\Psi_g(t))^2}=\expval{(\int_\Omega dx\,\Phi(x,t)g(x))^2}$, where $g(x)\in\mathcal{L}_2(\Omega)$ as well. As is the case in e.g. equation \eqref{eq:ExpvalHkSquaredGen}, the second moment is determined by the Fourier-coefficients $\Phi_k(t)$ of $\Phi(x,t)$ and $g_k$ of $g(x)$, namely
\begin{equation}
\expval{\left(\Psi_g(t)\right)^2}=\sum_{k,l}\cc{g_k}\,\cc{g_l}\expval{\Phi_k(t)\Phi_l(t)}.\label{eq:Equiv1}
\end{equation}
Like the KPZ equation, \eqref{eq:GeneralFTLangevin} is driven by spatially homogeneous Gaussian white noise $\eta(x,t)$ with two-point correlations of the Fourier-coefficients $\eta_k(t)$ given by $\expval{\eta_k(t)\eta_l(t)}\sim\delta_{k,-l}$. Therefore, we expect the solution to \eqref{eq:GeneralFTLangevin} subject to periodic boundary conditions to be spatially homogeneous as well, at least in the steady state, which implies 
\begin{equation}
\expval{\Phi_k(t)\Phi_l(t)}\sim\delta_{k,-l},\label{eq:Equiv2}
\end{equation}
see e.g. \cite{Hayot1997,McComb1990Book}. Hence, with \eqref{eq:Equiv2}, the expression in \eqref{eq:Equiv1} becomes
\begin{equation}
\expval{\left(\Psi_g(t)\right)^2}=g_0^2\expval{\left(\Phi_0(t)\right)^2}+\sum_{k\neq0}|g_k|^2\expval{\Phi_k(t)\Phi_{-k}(t)}.\label{eq:Equiv3}
\end{equation}
Comparing \eqref{eq:Equiv3} to $\expval{\normZero{\Phi(x,t)}^2}$, which is given by
\begin{equation}
\expval{\normZero{\Phi(x,t)}^2}=\sum_k\expval{\Phi_k(t)\Phi_{-k}(t)}=\expval{\left(\Phi_0(t)\right)^2}+\sum_{k\neq0}\expval{\Phi_k(t)\Phi_{-k}(t)},\label{eq:Equiv4}
\end{equation}
we find in the NESS
\begin{equation}
\expval{\left(\Psi_g(t)\right)^2}\simeq g_0^2\expval{\normZero{\Phi(x,t)}^2},\label{eq:Equiv5}
\end{equation}
provided that the long-time behavior is dominated by the Fourier-mode with largest eigenvalue, i.e. $k=0$ with $\mu_0=0$. Under the same condition, the first moment of the projected output reads in the NESS
\begin{equation}
\expval{\Psi_g(t)}=\sum_k\cc{g_k}\expval{\Phi_k(t)}\simeq g_0\expval{\Phi_0(t)},\label{eq:Equiv6}
\end{equation}
and thus
\begin{equation}
\left(\expval{\Psi_g(t)}\right)^2\simeq g_0^2\left(\expval{\Phi_0(t)}\right)^2.\label{eq:Equiv7}
\end{equation}
Similarly, 
\begin{equation}
\normZero{\expval{\Phi(x,t)}}^2=\sum_k\left|\expval{\Phi_k(t)}\right|^2\simeq\left(\expval{\Phi_0(t)}\right)^2\qquad\text{for }t\gg1,\label{eq:Equiv8}
\end{equation}
which implies
\begin{equation}
\left(\expval{\Psi_g(t)}\right)^2\simeq g_0^2\normZero{\expval{\Phi(x,t)}}^2.\label{eq:Equiv9}
\end{equation}
Note, that $g_0$ and $\Phi_0(t)$ have to be real throughout the argument (which is indeed the case for expansions with respect to the eigenfunctions of the general diffusion operators $\hat{L}$ from above). Hence, under the assumption that the prior mentioned requirements are met, which, of course, would have to be checked for every individual system (as was done in this section for the KPZ equation), the asymptotic equivalence in \eqref{eq:Equiv5} and \eqref{eq:Equiv9} validates the statement in \eqref{eq:EquivalenceProjectionNorm} (and therefore, in the NESS, also \eqref{eq:EpsilonSquaredNorm}) for a whole class of one-dimensional scalar SPDEs from \eqref{eq:GeneralFTLangevin}.

\subsection{Total Entropy Production for the KPZ Equation}\label{subsec:TotEntropyProductionKPZ}

The total entropy production for the KPZ equation is obtained by inserting $F_\gamma[h_\mu(\fr,t)]=\partial_x^2h(x,t)+\frac{\lambda_\text{eff}}{2}\left(\partial_xh(x,t)\right)^2$ and the explicit expression for the one-dimensional stationary probability distribution $p^s[h]$ into \eqref{eq:DefGenTotalEntropyProd}. The form of the latter is given in the following.

\paragraph{The Fokker--Planck Equation and its 1D Stationary Solution}\label{par:FokkerPlanck}

Let us briefly recapitulate the Fokker-Planck equation and its stationary solution in one spatial dimension for the KPZ equation.\\
The Fokker-Planck equation corresponding to \eqref{eq:RescaledKPZDimensionless} for the functional probability distribution $p[h]$ reads, e.g. \cite{HalpinHealyReview1995,AltlandBook2010,Frusawa2019,Takeuchi2017},
\begin{align}
\begin{split}
\parderiv{p[h]}{t}&=-\int_0^1dx\funcderiv{}{h}\left[\left(\partial_x^2h(x,t)+\frac{\lambda_\text{eff}}{2}(\partial_xh(x,t))^2\right)p[h]-\frac{1}{2}\funcderiv{p[h]}{h}\right]\\
&=-\int_0^1dx\funcderiv{j[h]}{h},
\end{split}\label{eq:DefFokkerPlanck}\\
j[h]&\equiv\left(\partial_x^2h(x,t)+\frac{\lambda_\text{eff}}{2}(\partial_xh(x,t))^2\right)p[h]-\frac{1}{2}\funcderiv{p[h]}{h},\label{eq:ProbCurrent}
\end{align}
with $j[h]$ as a probability current.\\
It is well known that for the case of pure Gaussian white noise, a stationary solution, i.e. $\partial_tp^s[h]=0$, to the Fokker-Planck equation is given by \cite{HalpinHealyReview1995,KrugReview1997,Takeuchi2017}
\begin{equation}
p^s[h]\equiv\exp\left[-\normZero{\partial_xh(x,t)}^2\right].\label{eq:WhiteNoiseStatProb}
\end{equation}
This stationary solution is the same as the one for the linear case, namely for the Edwards-Wilkinson model. Note that in \eqref{eq:WhiteNoiseStatProb} we denote by $\normZero{\cdot}^2$ the standard $\mathcal{L}_2$-norm.

\paragraph{Stationary Total Entropy Production}\label{par:StationaryTotEntropyProd}

With \eqref{eq:DefGenTotalEntropyProd}, the total entropy production in the NESS for the KPZ equation reads
\begin{align}
\begin{split}
\Delta s_\text{tot}&=\Delta s_m+s^\text{stat}=2\int_0^t dt^\prime\left(\dot{h},\left[\partial_x^2h+\frac{\lambda_\text{eff}}{2}(\partial_xh)^2\right]\right)_0-\left(h,\partial_x^2h\right)_0\\
&=\left[2\int_0^t dt^\prime\left(\dot{h},\partial_x^2h\right)_0-\left(h,\partial_x^2h\right)_0\right]+\lambda_\text{eff}\int_0^t dt^\prime\left(\dot{h},(\partial_xh)^2\right)_0.
\end{split}\label{eq:DeltaSTotStatInterm1}
\end{align}
Using $\left(\dot{h},\partial_x^2h\right)_0=\frac{1}{2}\deriv{}{t}\left(h,\partial_x^2h\right)_0$, and the initial condition $h(x,0)=0$, the first term in \eqref{eq:DeltaSTotStatInterm1} vanishes and thus 
\begin{equation}
\Delta s_\text{tot}=\lambda_\text{eff}\int_0^t dt^\prime\left(\dot{h}(x,t^\prime),\left(\partial_xh(x,t^\prime)\right)^2\right)_0.\label{eq:StatTotEntropyProductionKPZ}
\end{equation}
For Gaussian white noise, the expectation value of \eqref{eq:StatTotEntropyProductionKPZ} is given by 
\begin{align}
\begin{split}
\expval{\Delta s_\text{tot}}&=\lambda_\text{eff}\int_0^t dt^\prime\expval{\left(\dot{h}(x,t^\prime),\left(\partial_xh(x,t^\prime)\right)^2\right)_0}\\
&=\frac{\lambda_\text{eff}^2}{2}\int_0^tdt^\prime\expval{\normZero{\left(\partial_xh(x,t^\prime)\right)^2}^2}.
\end{split}\label{eq:ExpValStatTotEntropyProductionKPZ}
\end{align}
For a derivation of this result see \autoref{app:ExpValTotEntProd}. Note that \eqref{eq:ExpValStatTotEntropyProductionKPZ} and its derivation remains true for $h\in\text{span}\{\phi_{-\Lambda},\,\ldots,\,\phi_\Lambda\}$. More generally, the expectation of the total entropy production may also be written as
\begin{equation}
\expval{\Delta s_\text{tot}}=\frac{\lambda_\text{eff}^2}{2}\int_0^tdt^\prime\,\expval{\left(\left(\partial_xh(x,t^\prime)\right)^2,\hat{K}^{-1}\left(\partial_xh(x,t^\prime)\right)^2\right)_0},\label{eq:ExpValStatTotEntropyProductionKPZColoredNoise}
\end{equation}
with $\hat{K}^{-1}$ from \eqref{eq:DefInvNoiseOperatorScaled}.

\paragraph{Evaluating the Expectation of the Stationary Total Entropy Production}\label{par:EvalStatTotEntProd}
Above, an expression for the stationary total entropy production $\Delta s_\text{tot}$ and its expectation value were derived (see eq. \eqref{eq:ExpValStatTotEntropyProductionKPZ}). Inserting the Fourier representation from \eqref{eq:GenEigFuncExpansion} and \eqref{eq:DefEigenFunc} into \eqref{eq:ExpValStatTotEntropyProductionKPZ} leads to
\begin{align}
\begin{split}
&\expval{\Delta s_\text{tot}}\\
&=(4\pi^2)^2\frac{\lambda_\text{eff}^2}{2}\int_0^tdt^\prime\int_0^1dx\sum_{\inR{k}}\sum_{\inR{m}}e^{2\pi ix(k-m)}\\
&\quad\times\expval{\sum_{\inRkWO{l}{k}{,k}}l(k-l)h_l(t^\prime)h_{k-l}(t^\prime)\sum_{\inRkWO{n}{m}{,m}}n(m-n)\cc{h_n}(t^\prime)\cc{h_{m-n}}(t^\prime)}\\
&=(4\pi^2)^2\frac{\lambda_\text{eff}^2}{2}\int_0^tdt^\prime\sum_{\inR{k}}\;\sum_{\inRkWO{l,n}{k}{,k}}l(k-l)n(k-n)\\
&\quad\times\expval{h_l(t^\prime)h_{k-l}(t^\prime)\cc{h_n}(t^\prime)\cc{h_{k-n}}(t^\prime)},
\end{split}\label{eq:ExpValDeltaSTotStatInterm1}
\end{align}
with $\mathfrak{R}_k$ from \eqref{eq:DefRk}. As \eqref{eq:ExpValDeltaSTotStatInterm1} above is already of order $\lambda_\text{eff}^2$, it suffices to expand the Fourier coefficients $h_i(t^\prime)$ to zeroth order, which yields
\begin{align}
\begin{split}
\expval{\Delta s_\text{tot}}&=(4\pi^2)^2\frac{\lambda_\text{eff}^2}{2}\int_0^tdt^\prime\sum_{\inR{k}}\;\sum_{\inRkWO{l,n}{k}{,k}}l(k-l)n(k-n)\\
&\qquad\times\expval{\hh{l}{0}(t^\prime)\hh{k-l}{0}(t^\prime)\cc{\hh{n}{0}}(t^\prime)\cc{\hh{k-n}{0}}(t^\prime)}+O(\lambda_\text{eff}^4),
\end{split}\label{eq:ExpValDeltaSTotStatInterm2}
\end{align}
with $\hh{i}{0}(t^\prime)$ given by \eqref{eq:DefHk0}. Via a Wick contraction and using \eqref{eq:DefCorrFuncHks}, the four-point correlation function in \eqref{eq:ExpValDeltaSTotStatInterm2} reads
\begin{align}
\begin{split}
&\expval{\hh{l}{0}(t^\prime)\hh{k-l}{0}(t^\prime)\cc{\hh{n}{0}}(t^\prime)\cc{\hh{k-n}{0}}(t^\prime)}\\
&=\Pi_{l,k-l}(t^\prime,t^\prime)\Pi_{-n,n-k}(t^\prime,t^\prime)\delta_{0,k}+\Pi_{l,-n}(t^\prime,t^\prime)\Pi_{k-l,n-k}(t^\prime,t^\prime)\delta_{l,n}\\
&\qquad+\Pi_{l,n-k}(t^\prime,t^\prime)\Pi_{k-l,-n}(t^\prime,t^\prime)\delta_{n,k-l}.
\end{split}\label{eq:ExpValDeltaSTotStatWickContraction}
\end{align}
Inserting \eqref{eq:ExpValDeltaSTotStatWickContraction} into \eqref{eq:ExpValDeltaSTotStatInterm2} leads to the following form of the total entropy production in the NESS,
\begin{align}
\begin{split}
&\expval{\Delta s_\text{tot}}\\
&=(4\pi^2)^2\frac{\lambda_\text{eff}^2}{2}\left[\sum_{\inRWO{l,n}{}}\frac{l^2n^2}{4\mu_l\mu_n}+2\sum_{\inR{k}}\;\sum_{\inRkWO{l}{k}{,k}}\frac{l^2(k-l)^2}{4\mu_l\mu_{k-l}}\right]\,t+O(\lambda_\text{eff}^4).
\end{split} \label{eq:StatTotEntropyProductionKPZFinal}
\end{align}
Note that the long time behavior of $\expval{\Delta s_\text{tot}}$ is indeed of the form required, i.e. $\expval{\Delta s_\text{tot}}\sim\lambda_\text{eff}^2t$ (see remark after \eqref{eq:EpsilonSquaredKPZ}), for the uncertainty relation to hold. With $\mu_k$ from \eqref{eq:EigenvaluesOfL}, the expression for the total entropy production from \eqref{eq:StatTotEntropyProductionKPZFinal} reads
\begin{equation}
\expval{\Delta s_\text{tot}}=\frac{\lambda_\text{eff}^2}{2}\left[\Lambda^2+\frac{3\Lambda^2-\Lambda}{2}\right]t+O(\lambda_\text{eff}^4).\label{eq:StatTotEntropyProductionKPZFinalSimplified}
\end{equation}
Thus, with \eqref{eq:DefGenSigma} and \eqref{eq:StatTotEntropyProductionKPZFinalSimplified}, the total entropy production rate reads
\begin{equation}
\sigma=\frac{\lambda_\text{eff}^2}{2}\left[\Lambda^2+\frac{3\Lambda^2-\Lambda}{2}\right]+O(\lambda_\text{eff}^4).\label{eq:SigmaKPZ}
\end{equation}\par
With \eqref{eq:EpsilonSquaredKPZ} and \eqref{eq:StatTotEntropyProductionKPZFinalSimplified}, or, equivalently, \eqref{eq:StatCurrentKPZ}, \eqref{eq:DiffusivityKPZ} and \eqref{eq:SigmaKPZ}, the constituents of the thermodynamic uncertainty relation are known. Hence, the product entering the TUR from \eqref{eq:DefGenFTTUR} for the KPZ equation reads
\begin{equation}
\expval{\Delta s_\text{tot}}\,\epsilon^2=\frac{2\sigma\,D_g}{J_g^2}=2+\left(3-\frac{1}{\Lambda}\right)+O(\lambda_\text{eff}^2).\label{eq:TURKPZFinal}
\end{equation}
Here, we deliberately refrain from writing $\expval{\Delta s_\text{tot}}\,\epsilon^2=5-1/\Lambda$ as this would somewhat mask the physics causing this result. This point will be discussed further in the following.

\paragraph{Edwards--Wilkinson Model for a Constant Driving Force}\label{par:EdwardsWilkinson}
To give an interpretation of the two terms in \eqref{eq:StatTotEntropyProductionKPZFinal} and consequently in \eqref{eq:TURKPZFinal}, we believe it instructive to briefly calculate the precision and total entropy production for the case of the one-dimensional Edwards-Wilkinson model modified by an additional constant non-random driving \lq force\rq{} $v_0$ and subject to periodic boundary conditions. To be specific, we consider
\begin{equation}
\partial_th(x,t)=\partial_x^2h(x,t)+v_0+\eta(x,t)\qquad x\in[0,1],\label{eq:EdwardsWilkinsonEq}
\end{equation}
already in dimensionless form and with space-time white noise $\eta$. We denote \eqref{eq:EdwardsWilkinsonEq} in the sequel with FEW for \lq forced Edwards-Wilkinson equation\rq{}. Following the same procedure as described in \autoref{sec:TheoBack}, we find the following integral equation for the $k$-th Fourier coefficient of the height field in FEW,
\begin{equation}
h_k(t)=e^{\mu_kt}\int_0^tdt^\prime\,e^{-\mu_kt^\prime}\left[v_0\delta_{0,k}+\eta_k(t^\prime)\right],\label{eq:EWHk}
\end{equation}
where again a flat initial configuration was assumed and $\mu_k=-4\pi^2k^2$ as above. With \eqref{eq:EWHk}, we get immediately in the NESS
\begin{equation}
\expval{\Psi_g(t)}=g_0v_0t=J_gt,\label{eq:EWExpValH}
\end{equation}
and thus $\expval{\Psi_g(t)}^2=g_0^2v_0^2t^2$ as well as
\begin{align*}
&\expval{\left(\Psi_g(t)\right)^2}\\
&=\sum_{\inR{k,l}}\cc{g_k}\,\cc{g_l}\expval{h_k(t)h_l(t)}\\
&=\sum_{\inR{k,l}}\cc{g_k}\,\cc{g_l}e^{(\mu_k+\mu_l)t}\int_0^tdr\int_0^tds\,e^{-\mu_kr-\mu_ls}\left(v_0^2\delta_{0,k}\delta_{0,l}+\expval{\eta_k(r)\eta_l(s)}\right)\\
&=g_0^2v_0^2t^2+\sum_{\inR{k}}|g_k|^2\frac{e^{2\mu_kt}-1}{2\mu_k}\\
&=g_0^2v_0^2t^2+g_0^2t,\qquad\text{for }t\gg1.
\end{align*}
Thus,
\begin{equation}
\epsilon^2=\frac{\expval{\left(\Psi_g(t)-\expval{\Psi_g(t)}\right)^2}}{\expval{\Psi_g(t)}^2}\simeq\frac{t}{v_0^2t^2}=\frac{1}{v_0^2}\frac{1}{t}.\label{eq:EWPrecision}
\end{equation}
As already discussed above in \autoref{sec:TheoBack}, the Fokker-Planck equation corresponding to \eqref{eq:EdwardsWilkinsonEq} has the stationary solution $p^s[h]=\exp\left[-\int dx\,(\partial_xh)^2\right]$ and thus, with \eqref{eq:DefGenTotalEntropyProd} and \eqref{eq:EWHk}, the total entropy production reads in the NESS
\begin{equation}
\expval{\Delta s_\text{tot}}=2\int_0^1dx\,\expval{h(x,t)}\,v_0=2\,v_0^2\,t.\label{eq:EWTotEntProd}
\end{equation}
With \eqref{eq:EWPrecision} and \eqref{eq:EWTotEntProd}, the TUR product for \eqref{eq:EdwardsWilkinsonEq} is given by
\begin{equation}
\expval{\Delta s_\text{tot}}\,\epsilon^2=2,\label{eq:EWTUR}
\end{equation}
i.e. the thermodynamic uncertainty relation is indeed saturated for the Edwards-Wilkinson equation subject to a constant driving \lq force\rq{} $v_0$. For the sake of completeness we state the expressions for the current, diffusivity and rate of entropy production in the non-equilibrium steady state, namely
\begin{equation}
J_g^\text{FEW}=g_0v_0,\qquad D_g^\text{FEW}=\frac{g_0^2}{2},\qquad\sigma^\text{FEW}=2\,v_0^2.\label{eq:EWJDSigma}
\end{equation}\par
With the calculations for FEW, we can now give an interpretation of the two terms in \eqref{eq:StatTotEntropyProductionKPZFinalSimplified} and \eqref{eq:TURKPZFinal}. The first term in squared brackets in \eqref{eq:StatTotEntropyProductionKPZFinalSimplified} originates from the first term of \eqref{eq:StatTotEntropyProductionKPZFinal}, where the latter represents the action of all higher-order Fourier modes on the mode $k=0$ (see \eqref{eq:ExpValDeltaSTotStatWickContraction}). To illustrate this point further, observe that, in the NESS, we get according to \eqref{eq:MeanSquaredGen} to \eqref{eq:StatCurrentKPZ} for the current:
\begin{equation}
J_g=2\pi^2g_0\lambda_\text{eff}\left(\sum_{\inRwo{l}}\frac{l^2}{2(-\mu_l)}\right)=g_0\frac{\lambda_\text{eff}}{2}\Lambda,\label{eq:NESSCurrent}
\end{equation}
and from the calculation above we see that it contains only the impact of Fourier modes $l\neq0$ on the mode $k=0$, which belongs to the constant eigenfunction $\phi_0(x)=1$. In other words, the modes $l\neq0$ act like a constant external excitation, just in the same manner as $v_0$ acts for FEW in \eqref{eq:EWExpValH}. Comparing \eqref{eq:NESSCurrent} to \eqref{eq:EWExpValH}, we may set
\begin{equation}
v_0=2\pi^2\lambda_\text{eff}\left(\sum_{\inRwo{l}}\frac{l^2}{2(-\mu_l)}\right)=\frac{\lambda_\text{eff}}{2}\Lambda,\label{eq:RelationVoToLambda}
\end{equation}
and get $J_g=g_0v_0$ in both cases.\\
Following now the calculations for FEW, we would expect from \eqref{eq:EWTotEntProd}
\begin{equation}
\expval{\Delta s_\text{tot}}=2v_0^2\,t=(4\pi^2)^2\frac{\lambda_\text{eff}^2}{2}\left(\sum_{\inRwo{l}}\frac{l^2}{2(-\mu_l)}\right)^2\,t=\frac{\lambda_\text{eff}^2}{2}\Lambda^2\,t,\label{eq:DeltaSTot}
\end{equation}
which is in fact exactly the first term in the squared brackets from \eqref{eq:StatTotEntropyProductionKPZFinal} and \eqref{eq:StatTotEntropyProductionKPZFinalSimplified}, respectively. Since with \eqref{eq:RelationVoToLambda} also the expression for $\epsilon^2$ from \eqref{eq:EWPrecision} coincides with the first summand on the r.h.s. of \eqref{eq:EpsilonSquaredKPZ}, it is clear that both cases result in the saturated TUR. This explains the value $2$ on the r.h.s. of \eqref{eq:TURKPZFinal}.\\
Turning to the second term of \eqref{eq:StatTotEntropyProductionKPZFinalSimplified}, we see that it stems from the second term in \eqref{eq:StatTotEntropyProductionKPZFinal}. In contrast to the first term in \eqref{eq:StatTotEntropyProductionKPZFinal}, the second one does not only measure the effect of the modes on the $k=0$ mode but also on all other modes $k\neq0$. It further features interactions of the $k$ and $l$ modes among each other via mode coupling. Hence, the mode coupling seems responsible for the larger constant on the right hand side of \eqref{eq:TURKPZFinal}, since by neglecting the mode coupling term in \eqref{eq:StatTotEntropyProductionKPZFinalSimplified}, the thermodynamic uncertainty relation was saturated also for the KPZ equation up to $O(\lambda_\text{eff}^2)$. To conclude this brief discussion, we give the respective relations of the KPZ current \eqref{eq:StatCurrentKPZ}, diffusivity \eqref{eq:DiffusivityKPZ} and total entropy production rate \eqref{eq:SigmaKPZ} to FEW, namely
\begin{align}
\begin{split}
J_g^\text{KPZ}&=J_g^\text{FEW}+O(\lambda_\text{eff}^3),\\
D_g^\text{KPZ}&=D_g^\text{FEW}+g_0^2\frac{\lambda_\text{eff}^2}{64\,\pi^2}\mathcal{H}_\Lambda^{(2)}+O(\lambda_\text{eff}^4),\\
\sigma^\text{KPZ}&=\sigma^\text{FEW}+\lambda_\text{eff}^2\frac{3\Lambda^2-\Lambda}{4}+O(\lambda_\text{eff}^4),
\end{split}\label{eq:RelationJDSigmaKPZToEW}
\end{align}
with $J_g^\text{FEW}$, $D_g^\text{FEW}$ and $\sigma^\text{FEW}$ from \eqref{eq:EWJDSigma}. We see that the additional mode coupling term in KPZ leads to corrections in $D_\text{KPZ}$ and $\sigma_\text{KPZ}$ of at least second order in $\lambda_\text{eff}$. For the case of $\lambda_\text{eff}\to0$ the KPZ equation becomes the standard Edwards-Wilkinson equation (EW), namely $\partial_th(x,t)=\partial_x^2h(x,t)+\eta(x,t)$, which possesses a genuine equilibrium steady state. Therefore, for the standard EW we have $J_g^\text{EW}=0$, $\sigma^\text{EW}=0$ and $D_g^\text{EW}=g_0^2/2$. From \eqref{eq:RelationJDSigmaKPZToEW} it follows that for $\lambda_\text{eff}\to0$, $(J_g,\sigma,D_g)_\text{KPZ}\to(J_g,\sigma,D_g)_\text{FEW}$ and from \eqref{eq:EWJDSigma}, \eqref{eq:RelationVoToLambda} that $(J_g,\sigma,D_g)_\text{FEW}\to(J_g,\sigma,D_g)_\text{EW}=(0,0,g_0^2/2)$. Hence, the non-zero expressions for $J_g^\text{KPZ}$ and $\sigma^\text{KPZ}$ result solely from the KPZ non-linearity. The impact of the latter on the $k=0$ Fourier mode (i.e. the spatially constant mode) results in contributions to $J_g^\text{KPZ}$ and $\sigma^\text{KPZ}$ that can be modeled exactly by FEW, the Edwards-Wilkinson equation driven by a constant force $v_0$ from \eqref{eq:EdwardsWilkinsonEq}.

\section{Conclusion}\label{sec:Conclusion}

We have introduced an analog of the TUR \cite{BaratoSeifertUR2015,Gingrich2016}  in a general field-theoretic setting (see \eqref{eq:DefGenFTTUR}) and shown its validity for the Kardar-Parisi-Zhang equation up to second order of perturbation. To ensure convergence of the quantities entering the thermodynamic uncertainty relation for the case of Gaussian space-time white noise, we had to introduce an arbitrarily large but finite cutoff $\Lambda$ of the corresponding Fourier spectrum. While this cutoff solves the issue of divergences, it naturally leads to subtleties in treating the non-linearity, as its Fourier spectrum is affected also by modes that are beyond the considered spectral range. In order to minimize the resulting bias, the cutoff has to be chosen large enough such as to guarantee the dominance of the diffusive term over the non-linear term. To circumvent the introduction of a cutoff to ensure convergence, a possible solution may be to induce a higher regularity by treating spatially colored noise instead of Gaussian white noise and/or choosing a higher order diffusion operator $\hat{L}$ (see e.g. \cite{WangXu2010,Wolf1990}) . This is currently under investigation.\par 
As is obvious from \eqref{eq:TURKPZFinal}, the field-theoretic version of the TUR for the KPZ equation displays a greater constant than the one in \cite{BaratoSeifertUR2015}. This is due to the mode-coupling of the fields as a consequence of the KPZ non-linearity. To illustrate this point, we also treated the Edwards-Wilkinson equation in \autoref{par:EdwardsWilkinson}, driven out of equilibrium by a constant velocity $v_0$, see \eqref{eq:EdwardsWilkinsonEq}. By identifying $v_0$ with the influence of higher-order Fourier modes on the mode $k=0$, we may interpret the first term in \eqref{eq:StatTotEntropyProductionKPZFinal} as the contribution from the forced Edwards-Wilkinson equation, for which the TUR with constant equal to $2$ is saturated (see \eqref{eq:EWTUR}), an observation which is in accordance with findings in \cite{Gingrich2017} for finite dimensional driven diffusive systems. The second term in \eqref{eq:StatTotEntropyProductionKPZFinal} is the contribution to the entropy production made up by the interaction between Fourier modes of arbitrary order, which is due to the mode coupling generated by the KPZ non-linearity. It is this additional entropy production that weakens the dissipation bound in the TUR. Note, that also the first term in \eqref{eq:StatTotEntropyProductionKPZFinal} is due to the mode-coupling, however is special in thus far that it measures only the impact of the other modes on the zeroth $k$-mode and does not include a response of the mode $k=0$.\par
Regarding future research, an intriguing topic is the question as to whether the findings in \cite{Gingrich2017} concerning conditions for the saturation of the dissipation bound in the TUR for an overdamped two-dimensional Langevin equation can be recovered in the present field-theoretic setting. Furthermore, it would be of great interest to employ the developed framework to other field-theoretic Langevin equations in order to observe the resulting dissipation bounds in the corresponding TURs. Of special interest in this context is the stochastic Burgers equation, especially, if excited by a noise term suitable for generating genuine turbulent response (see \cite{Yakhot1995}). A comparison of the predictions made in the present paper to numerical simulations of the KPZ equation seems to be another intriguing task. Besides numerical calculations, it would also be of great interest to test our predictions via experimental realizations of KPZ interfaces. Lastly, the formulation of a genuine non-perturbative, analytic formalism would also be of utmost interest.

\vfill
\appendix

\section{Evaluation of \eqref{eq:ExpvalHkSquaredGen}}\label{app:EvaluationExpValHkSquared}

Using \eqref{eq:DefCorrFuncHks}, the first term in \eqref{eq:ExpvalHkSquaredGen} reads
\begin{equation}
\expval{\hh{k}{0}(t)\hh{l}{0}(t)}=\Pi_{k,l}(t,t)\delta_{k,-l}=e^{(\mu_k+\mu_l)t}\frac{1-e^{-(\mu_k+\mu_l)t}}{\mu_k+\mu_l}\delta_{k,-l}.\label{eq:ExpValHk0Hl0}
\end{equation}
Note that the case of $k=0$ is treated like in \eqref{eq:ExpValHk1Interm1}. The second term in \eqref{eq:ExpvalHkSquaredGen} is given by
\begin{align}
\begin{split}
&\expval{\hh{k}{1}(t)\hh{l}{1}(t)}\\
&=(2\pi^2)^2\sum_{\inRkWO{m}{k}{,k}}m(k-m)\sum_{\inRkWO{n}{l}{,l}}n(l-n)\int_0^tdt^\prime\,e^{\mu_k(t-t^\prime)}\int_0^tdr\,e^{\mu_l(t-r)}\\
&\qquad\times\expval{\hh{m}{0}(t^\prime)\hh{k-m}{0}(t^\prime)\hh{n}{0}(r)\hh{l-n}{0}(r)}\\
&=-2(2\pi^2)^2\sum_{\inRkWO{m}{k}{,k}}m^2(k-m)(l+m)\int_0^tdt^\prime\,e^{\mu_k(t-t^\prime)}\int_0^tdr\,e^{\mu_l(t-r)}\\
&\qquad\times\Pi_{m,m}(t^\prime,r)\Pi_{k-m,l+m}(t^\prime,r)\delta_{k,-l}\\
&\quad+(2\pi^2)^2\sum_{\inRkWO{m}{k}{,k}}m(k-m)\sum_{\inRkWO{n}{l}{,l}}n(l-n)\int_0^tdt^\prime\,e^{\mu_k(t-t^\prime)}\int_0^tdr\,e^{\mu_l(t-r)}\\
&\qquad\times\Pi_{m,k-m}(t^\prime,t^\prime)\Pi_{n,l-n}(r,r)\delta_{0,k}\delta_{0,l},\label{eq:ExpValHk1Hl1}
\end{split}
\end{align}
where we used Wick's-theorem, \eqref{eq:DefCorrFuncHks} and \eqref{eq:DefHk1}. Note that the two Kronecker-deltas in the last term of \eqref{eq:ExpValHk1Hl1} can also be written as $\delta_{0,k}\delta_{0,l}\delta_{k,-l}$, such that the whole expression is multiplied by $\delta_{k,-l}$. Again with Wick's-theorem, \eqref{eq:DefCorrFuncHks} and \eqref{eq:DefHk2} we can calculate the third and forth term of \eqref{eq:ExpvalHkSquaredGen} accordingly and find
\begin{align}
\begin{split}
&\expval{\hh{k}{0}(t)\hh{l}{2}(t)}\\
&=4(2\pi^2)^2\sum_{\inRkWO{m}{l}{,l}}ml(l-m)(m-l)\int_0^tdt^\prime\,e^{\mu_l(t-t^\prime)}\int_0^{t^\prime}dr\,e^{\mu_m(t^\prime-r)}\\
&\qquad\times\Pi_{k,l}(t,r)\Pi_{l-m,l-m}(t^\prime,r)\delta_{k,-l},\\\\
&\expval{\hh{k}{2}(t)\hh{l}{0}(t)}\\
&=4(2\pi^2)^2\sum_{\inRkWO{m}{k}{,k}}mk(k-m)(m-k)\int_0^tdt^\prime\,e^{\mu_k(t-t^\prime)}\int_0^{t^\prime}dr\,e^{\mu_m(t^\prime-r)}\\
&\qquad\times\Pi_{k,l}(t,r)\Pi_{k-m,k-m}(t^\prime,r)\delta_{k,-l}.
\end{split}\label{eq:ExpValHk0Hl2}
\end{align}
As can be seen from \eqref{eq:ExpValHk0Hl0} to \eqref{eq:ExpValHk0Hl2}, all four terms in \eqref{eq:ExpvalHkSquaredGen} contain a $\delta_{k,-l}$ and thus \eqref{eq:ExpvalHkSquaredGen} reduces to
\begin{align}
\begin{split}
\expval{\left(\Psi_g(t)\right)^2}&=\sum_{\inR{k}}|g_k|^2\left[\expval{\hh{k}{0}(t)\hh{-k}{0}(t)}+\lambda_\text{eff}^2\left(\expval{\hh{k}{1}(t)\hh{-k}{1}(t)}\right.\right.\\
&\qquad\left.\left.+\expval{\hh{k}{0}(t)\hh{-k}{2}(t)}+\expval{\hh{k}{2}(t)\hh{-k}{0}(t)}\right)+O(\lambda_\text{eff}^4)\right]
\end{split}\label{eq:ExpvalHkSquaredGen1}
\end{align}
The first term of \eqref{eq:ExpvalHkSquaredGen1} is readily evaluated with \eqref{eq:ExpValHk0Hl0} as
\begin{equation}
\expval{\hh{k}{0}(t)\hh{-k}{0}(t)}=\Pi_{k,-k}(t,t)\delta_{k,k}=\frac{e^{2\mu_kt}-1}{2\mu_k}=\begin{cases}\quad t\qquad&\text{for}\;k=0, \\ -\frac{1}{2\mu_k}\qquad&\text{for}\;k\neq0\;\text{and}\;t\gg1.\end{cases}\label{eq:ExpValHk0Hk0}
\end{equation}
The second term of \eqref{eq:ExpvalHkSquaredGen1} reads with \eqref{eq:ExpValHk1Hl1}:
\begin{align}
\begin{split}
&\expval{\hh{k}{1}(t)\hh{-k}{1}(t)}\\
&=2(2\pi^2)^2\sum_{\inRkWO{m}{k}{,k}}m^2(k-m)^2\int_0^tdt^\prime\,e^{\mu_k(t-t^\prime)}\int_0^tdr\,e^{\mu_k(t-r)}\\
&\qquad\times\Pi_{m,m}(t^\prime,r)\Pi_{k-m,k-m}(t^\prime,r)\\
&\quad+(2\pi^2)^2\sum_{\inRwo{m}}m^2\sum_{\inRwo{n}}n^2\int_0^tdt^\prime\,\Pi_{m,m}(t^\prime,t^\prime)\int_0^tdr\,\Pi_{n,n}(r,r)
\end{split}\label{eq:ExpValHk1Hk1Interm1}
\end{align}
Hence, with $\expval{\hh{k}{1}(t)}$ from \eqref{eq:ExpValHk1Interm1}, the expression in \eqref{eq:ExpValHk1Hk1Interm1} becomes
\begin{align}
\begin{split}
&\expval{\hh{k}{1}(t)\hh{-k}{1}(t)}\\
&=2(2\pi^2)^2e^{2\mu_kt}\sum_{\inRkWO{m}{k}{,k}}\frac{m^2(k-m)^2}{4\mu_m\mu_{k-m}}\int_0^tdt^\prime\int_0^tdr\,e^{(\mu_m+\mu_{k-m}-\mu_k)(t^\prime+r)}\\
&\qquad\times\left(1-e^{-2\mu_m(t^\prime\wedge r)}-e^{-2\mu_{k-m}(t^\prime\wedge r)}+e^{-2(\mu_m+\mu_{k-m})(t^\prime\wedge r)}\right)+\expval{\hh{k}{1}(t)}^2.
\end{split}\label{eq:ExpValHk1Hk1Interm2}
\end{align}
Here the choice of the minimum of $(t^\prime\wedge r)$ is arbitrary, since for $(t^\prime\wedge r)=r$ the other case is obtained by simply interchanging $r\leftrightarrow t^\prime$ under the integral and vice versa; thus the results for both choices are equivalent. In the following $(t^\prime\wedge r)=r$ is chosen. Hence, the integral expression in \eqref{eq:ExpValHk1Hk1Interm2} can be evaluated as
\begin{align}
\begin{split}
&e^{2\mu_kt}\int_0^tdt^\prime\,e^{(\mu_m+\mu_{k-m}-\mu_k)t^\prime}\int_0^{t^\prime}dr\\
&\quad\times\left(e^{(\mu_m+\mu_{k-m}-\mu_k)r}-e^{(-\mu_m+\mu_{k-m}-\mu_k)r}-e^{(\mu_m-\mu_{k-m}-\mu_k)r}+e^{-(\mu_m+\mu_{k-m}+\mu_k)r}\right)\\
&\simeq\begin{cases} -\frac{t}{2\mu_m}\quad&\text{for}\;k=0 \\ \frac{1}{2\mu_k(\mu_m+\mu_{k-m}+\mu_k)}\quad&\text{for}\;k\neq0 \end{cases}\quad\text{for}\;t\gg1.
\end{split}\label{eq:ExpValHk1Hk1Interm3}
\end{align}
Thus, with \eqref{eq:ExpValHk1Hk1Interm2} and \eqref{eq:ExpValHk1Hk1Interm3}, the long time behavior of $\expval{\hh{k}{1}(t)\hh{-k}{1}(t)}$ is given by
\begin{align}
\begin{split}
&\expval{\hh{k}{1}(t)\hh{-k}{1}(t)}\\
&\simeq\expval{\hh{k}{1}(t)}^2+2(2\pi^2)^2\times\begin{cases}\left[-\sum_{\inRWO{l}{}}\frac{l^4}{8\mu_l^3}\right]\,t\quad&\text{for}\;k=0 ,\\  \\ \sum_{\inRkWO{l}{k}{,k}}\frac{l^2(k-l)^2}{4\mu_l\mu_{k-l}}\frac{1}{2\mu_k(\mu_l+\mu_{k-l}+\mu_k)}\quad&\text{for}\;k\neq0 ,\end{cases}
\end{split}\label{eq:ExpValHk1Hk1Asympt}
\end{align}
where we changed $m\to l$. To save computational effort, rewrite the last two terms of \eqref{eq:ExpvalHkSquaredGen1} in the following way
\begin{equation}
\expval{\hh{k}{0}(t)\hh{-k}{2}(t)}+\expval{\hh{k}{2}(t)\hh{-k}{0}(t)}=2\,\text{Re}\left[\expval{\hh{k}{0}(t)\cc{\hh{k}{2}}(t)}\right].\label{eq:ExpValHk0Hk2RewrittenRealPart}
\end{equation}
Hence, it suffices to calculate one of the two expectation values. With \eqref{eq:ExpValHk0Hl2} we see that
\begin{align}
\begin{split}
&\expval{\hh{k}{0}(t)\cc{\hh{k}{2}}(t)}\\
&=-16\pi^4e^{\mu_kt}\int_0^tdt^\prime\,e^{-\mu_kt^\prime}\sum_{\inRkWO{m}{k}{,k}}km^2(k-m)e^{\mu_{k-m}t^\prime}\int_0^{t^\prime}dr\,e^{-\mu_{k-m}r}\\
&\qquad\times\Pi_{k,k}(t,r)\Pi_{m,m}(t^\prime,r),
\end{split}\label{eq:ExpValHk0Hk2Interm2}
\end{align}
where we substituted $m\to k-m$ and used the symmetry of $\Pi_{k,l}(t,t^\prime)$ from \eqref{eq:SymmetryPsis}. Note that for $k=0$, the above expression in \eqref{eq:ExpValHk0Hk2Interm2} vanishes. Thus in the following calculations $k\neq0$ is assumed. In this setting, \eqref{eq:ExpValHk0Hk2Interm2} reads with \eqref{eq:DefCorrFuncHks}
\begin{align}
\begin{split}
&\quad-16\pi^4e^{\mu_kt}\int_0^tdt^\prime\,e^{-\mu_kt^\prime}\sum_{\inRkWO{l}{k}{,k}}kl^2(k-l)e^{\mu_{k-l}t^\prime}\int_0^{t^\prime}dr\,e^{-\mu_{k-l}r}\Pi_{k,k}(t,r)\Pi_{l,l}(t^\prime,r)\\
&=-16\pi^4\sum_{\inRkWO{l}{k}{,k}}\frac{kl^2(k-l)}{4\mu_k\mu_l}\frac{1}{2\mu_k(\mu_l+\mu_k+\mu_{k-l})}\qquad\text{for}\;k\neq0\;\text{and}\;t\gg1,
\end{split}\label{eq:ExpValHk0Hk2Interm2.1}
\end{align}
where we changed summation index $m\to l$. Thus, with the results from \eqref{eq:ExpValHk0Hk0}, \eqref{eq:ExpValHk1Hk1Asympt}, \eqref{eq:ExpValHk0Hk2RewrittenRealPart} and \eqref{eq:ExpValHk0Hk2Interm2.1}, the expectation value of  \eqref{eq:ExpvalHkSquaredGen} reads in the long-time asymptotics
\begin{align}
\begin{split}
&\expval{\left(\Psi_g(t)\right)^2}\simeq g_0^2\left[1-2(2\pi^2)^2\lambda_\text{eff}^2\sum_{\inRWO{l}{}}\frac{l^4}{8\mu_l^3}\right]\,t+g_0^2\lambda_\text{eff}^2\sum_{\inR{k}}\expval{\hh{k}{1}(t)}^2+O(\lambda_\text{eff}^4).
\end{split}
\end{align}

\section{Expectation of the Total Entropy Production}\label{app:ExpValTotEntProd}

The Fokker-Planck equation for the KPZ equation from \eqref{eq:RescaledKPZDimensionless} reads, like in \autoref{par:FokkerPlanck},
\begin{equation}
\partial_t p[h]=-\int_0^1dx\,\funcderiv{}{h}\left[\left(\partial_x^2h+\frac{\lambda_\text{eff}}{2}\left(\partial_xh\right)^2\right)p[h]-\frac{1}{2}\funcderiv{p[h]}{h}\right].\label{eq:Fokker_Planck}
\end{equation}
Due to the conservation of probability, there is a current $j[h]$ given by
\begin{align}
j[h]&=\left(\partial_x^2h+\frac{\lambda_\text{eff}}{2}\left(\partial_xh\right)^2\right)p[h]-\frac{1}{2}\funcderiv{p[h]}{h}\nonumber\\
&=\left(\partial_x^2h+\frac{\lambda_\text{eff}}{2}\left(\partial_xh\right)^2\right)p[h]-\frac{1}{2}\funcderiv{\ln p[h]}{h}p[h]\nonumber\\
&=\left[\left(\partial_x^2h+\frac{\lambda_\text{eff}}{2}\left(\partial_xh\right)^2\right)-\frac{1}{2}\funcderiv{\ln p[h]}{h}\right]p[h]\label{eq:Current}\\
&\equiv v[h]p[h].
\end{align}
Following \cite{SeifertStochTherm2008}, expectation values of expressions like $\expval{\dot{h}\mathcal{G}[h]}$ are interpreted as
\begin{align}
\expval{\dot{h}\mathcal{G}[h]}&=\expval{v[h]\mathcal{G}[h]}=\funcint{h}v[h]\mathcal{G}[h]p[h]\\
\Leftrightarrow\quad\expval{\dot{h}\mathcal{G}[h]}&=\funcint{h}j[h]\mathcal{G}[h].\label{eq:Exp_Val}
\end{align}
Since the goal is to find an expression for the expectation value of the total entropy production in the stationary state, $\Delta s_\text{tot}$, it is useful to choose $p[h]$ as being the stationary solution $p^s[h]$ of the one-dimensional Fokker-Planck equation, which is given by
\begin{equation}
p^s[h]=\exp\left[-\normZero{\partial_xh}^2\right]\label{eq:Stationary_Probability}.
\end{equation}
Inserting this in \eqref{eq:Current} yields for the stationary probability current
\begin{equation}
j^s[h]=\frac{\lambda_\text{eff}}{2}(\partial_xh)^2p^s[h],\label{eq:Stationary_Current}
\end{equation}
where it was used that
\begin{equation}
\funcderiv{p^s[h]}{h}=2p^s[h]\partial_x^2h.
\end{equation}
Using the result from \eqref{eq:Exp_Val} and \eqref{eq:Stationary_Current} leads to
\begin{equation}
\expval{\dot{h}\mathcal{G}[h]}=\funcint{h}j^s[h]\mathcal{G}[h]=\frac{\lambda_\text{eff}}{2}\funcint{h}(\partial_xh)^2\mathcal{G}[h]p^s[h]=\frac{\lambda_\text{eff}}{2}\expval{(\partial_xh)^2\mathcal{G}[h]}.\label{eq:Exp_Val_Stat}
\end{equation}
Here it is understood that $\expval{\cdot}$ now denotes the expectation value with regard to the stationary distribution $p^s[h]$.\\
The total stationary entropy production $\Delta s_\text{tot}$ is given by (see \eqref{eq:StatTotEntropyProductionKPZ})
\begin{equation}
\Delta s_\text{tot}=\lambda_\text{eff}\int_0^tdt^\prime\int_0^1dx\,\dot{h}(x,t^\prime)(\partial_xh(x,t^\prime))^2.\label{eq:DeltaStot}
\end{equation}
Hence its expectation value reads
\begin{equation}
\expval{\Delta s_\text{tot}}=\lambda_\text{eff}\int_0^tdt^\prime\int_0^1dx\,\expval{\dot{h}(x,t^\prime)(\partial_xh(x,t^\prime))^2},
\end{equation}
which is evaluated with the aid of \eqref{eq:Exp_Val_Stat}:
\begin{equation}
\expval{\Delta s_\text{tot}}=\frac{\lambda_\text{eff}^2}{2}\int_0^tdt^\prime\expval{\normZero{\left(\partial_xh(x,t^\prime)\right)^2}^2}.\label{eq:Exp_Val_DeltaStot}
\end{equation}

\section{Regularity Results for the one-dimensional KPZ Equation}\label{app:MildSolution}

Dealing with the one-dimensional KPZ equation allows us to make use of the equivalence to the stochastic Burgers equation and adapt the regularity results for the latter from \cite{DaPrato1994,Bloemker2013,BloemkerJentzen2013,BloemkerKamrani2013,Goldys2005}. In \autoref{subsec:KPZinSpectralForm} and \autoref{subsec:CloserLookAtNoise}, we found that our operators $\hat{L}$ and $\hat{K}$ share the same set of eigenfunctions, which simplifies the results obtained by the authors of \cite{Bloemker2013,Goldys2005} to the following. Under the assumption that
\begin{equation}
\sum_{\inN{k}}k^{2\rho-2}(\alpha_k^\text{B})^2<\infty\qquad\text{for some}\;\rho>0,\label{eq:NoiseAssumptionBloemker}
\end{equation}
it is guaranteed almost surely that the solution $u(x,t)$ of the one-dimensional noisy Burgers equation $u\in\mathcal{C}([0,T],H)$, $T>0$, with $H=\mathcal{L}_2([0,1])$ or even $H=\mathcal{C}([0,1])$ and the spectral Galerkin approximation converges in $H$ to the solution $u$. Utilizing the mapping from KPZ to Burgers via $u(x,t)\equiv-\partial_xh(x,t)$, with $h$ solution to the KPZ equation, which implies 
\begin{equation}
\eta^\text{B}(x,t)=-\partial_x\eta^\text{KPZ}(x,t),\label{eq:RelationEtaBurgersToKPZ}
\end{equation}
and therefore 
\begin{equation}
\alpha_k^\text{B}\sim k\,\alpha_k^\text{KPZ},\label{eq:RelationAlphaKsBurgersKPZ}
\end{equation}
we get the following result for the 1d-KPZ equation:
\begin{equation}
\sum_{\inN{k}}k^\epsilon(\alpha_k^\text{KPZ})^2<\infty\quad(\epsilon=2\rho>0)\qquad\Rightarrow\qquad h\in\mathcal{C}\left([0,T],H^1([0,1])\right).\label{eq:BloemkerAssumptionKPZ}
\end{equation}
Here $H^1([0,1])$ denotes the Sobolev space of order one on $[0,1]$, i.e. $f\in H^1([0,1])$ $\Leftrightarrow$ $\Vert f\Vert_{\mathcal{L}_2([0,1])}<\infty$ and $\Vert f^\prime\Vert_{\mathcal{L}_2([0,1])}<\infty$, where $f^\prime$ is understood as the weak derivative of $f$. It holds that $H^1([0,1])\subset\mathcal{L}_2([0,1])$, which is what we used in \autoref{subsec:CloserLookAtNoise}.\\
As spatial white noise is excluded by \eqref{eq:BloemkerAssumptionKPZ}, we introduced a finite cutoff $\Lambda$ of the Fourier spectrum instead of using our approximation to the KPZ solution as a spectral Galerkin scheme and letting $\Lambda\to\infty$. However, for the KPZ equation driven by spatially colored noise satisfying \eqref{eq:BloemkerAssumptionKPZ} or even an adapted version of \eqref{eq:BloemkerAssumptionKPZ} to a higher order diffusion operator as defined in \autoref{subsec:KPZinSpectralForm} (see e.g. \cite{WangXu2010,Wolf1990}), in future work we want to derive a TUR taking the full Fourier spectrum into account.\\
We would like to conclude with the following remark. Since a couple of years, there exists a complete existence and regularity theory for the KPZ equation driven by space-time white noise introduced by Hairer \cite{Hairer2011} (see also \cite{Gubinelli2017,Cannizzaro2018} and for further reading on the so-called regularity structures developed in \cite{Hairer2011} see \cite{Hairer2014}). In \cite{Hairer2011} it is shown that the solutions of the KPZ equation with mollified noise converge after a suitable renormalization to the solution of the renormalized KPZ equation with space-time white noise, when removing the regularization. It is due to this renormalization procedure (where a divergent quantity needs to be subtracted) and the poor regularity of the solution, that at present it is not obvious to us how the method developed in \cite{Hairer2011} can be of use for constructing a TUR.

\bibliographystyle{spphys}
\bibliography{bibliography_list}

\end{document}